\documentclass[letterpaper,11pt,leqno]{article}
\PassOptionsToPackage{table}{xcolor}
\usepackage{paper}
\usepackage{graphicx}
\usepackage{booktabs}
\usepackage{amsmath}
\usepackage{array}
\usepackage{colortbl}
\definecolor{rowhighlight}{HTML}{FFF3E0}   
\definecolor{rowsig}{HTML}{FFE0B2}         
\definecolor{rowgood}{HTML}{E8F5E9}        
\definecolor{rowbad}{HTML}{FFEBEE}         
\definecolor{rowgray}{HTML}{F5F5F5}        
\newcommand{\safefig}[3]{%
  \IfFileExists{#1}{\includegraphics[#2]{#1}}{\fbox{\parbox{0.85\textwidth}{\centering\vspace{2em}[Figure pending: \texttt{#3}]\vspace{2em}}}}%
}

\hypersetup{pdftitle={Polymarket v1 Database: Trade-Tape Microstructure Across a Full Lifecycle}}

\newcommand{\bib}{paper.bib}
\newcommand{\figdir}{figures}

\begin{document}

\title{Polymarket-v1 Database}

\author{%
  Boka Qin\textsuperscript{*}
  \and
  Rui Yang\textsuperscript{\dagger}%
}

\date{June 2026}

\begin{titlepage}
\renewcommand{\thefootnote}{\fnsymbol{footnote}}
\maketitle
\footnotetext[1]{Corresponding author. Washington University in St.\ Louis. \texttt{bokaqin@wustl.edu}.}
\footnotetext[2]{Southwest University of Political Science and Law.
\texttt{quant@t17.capital}.}
\let\thefootnote\relax\footnotetext{\noindent Both authors are associated with Time Seventeen. The dataset is available at
\url{https://huggingface.co/datasets/TimeSeventeen/Polymarket-v1}.}

\begin{abstract}
We introduce the Polymarket-v1 Database: the complete on-chain trade archive of Polymarket's first-generation CTF Exchange on Polygon, spanning 2022-11-21 to 2026-04-28 and covering the full contract lifecycle from first settlement to natural termination. The dataset comprises 1.20 billion trade records across 1.30 million markets with \$61 billion in nominal volume. Its defining feature is all ground-truth aggressor direction derived from the blockchain settlement layer, a property unavailable in existing prediction market archives, which rely on heuristic inference.
We use this truth-aligned archive to benchmark standard microstructure tools and document three findings. First, the tick rule and bulk volume classification achieve near-random \emph{aggregate} accuracy (49.83\% and 50.51\%), but this masks a systematic, correctable price-level gradient driven by positive trade direction autocorrelation and concentrated market-making -- two structural features of prediction markets that violate the mean-reversion assumption embedded in classical classifiers. Second, these classification errors propagate into downstream metrics: inferred VPIN diverges substantially from ground-truth VPIN, and OFI estimates are directionally biased, with material consequences for Transaction Cost Analysis. Third, ground-truth microstructure quality predicts forecasting performance in ways that classification-based proxies cannot recover: True VPIN positively predicts Brier scores, while Gibbs spread negatively predicts them -- a selection effect reflecting that high-spread niche markets attract informed specialists rather than noise traders. Replacing ground-truth metrics with classified proxies attenuates both relationships, illustrating that measurement accuracy at the transaction level is a prerequisite for reliable inference about prediction market design and probability calibration. The dataset is publicly available.
\end{abstract}

\end{titlepage}

\section{Introduction}\label{s:introduction}
Prediction markets have emerged as foundational infrastructure for real-time probability estimation across politics, sports, and finance. Classic surveys document their forecasting accuracy and information aggregation properties \citep{WolfersZitzewitz2004,BergForsytheNelsonRietz2008}. However, their microstructure---how prices form, how liquidity evolves over time, how informed trading is identified, and how institutional changes impact market quality---has lacked systematic empirical evidence spanning a complete platform lifecycle. The root cause is a structural data gap: off-chain matching engines do not preserve permanent records, samples are often truncated, trade directions can only be inferred, and crucially, researchers never observe the true value. This makes the core assumption of informed trading theories impossible to test directly \citep{Kyle1985,GlostenMilgrom1985,OHara1995}.

To fill this empirical gap, we introduce and release the \textbf{Polymarket-v1 Database}, the complete on-chain trade archive of Polymarket's first-generation CTF Exchange from its first trade on 2022-11-21 to its last settlement on 2026-04-28. The dataset covers 1.2 billion trades, 1.3 million markets, and \$61 billion in nominal volume. Crucially, each trade carries a ground-truth buyer/seller direction derived from the blockchain settlement layer, making it the first publicly available archive of its scale with a verified trade-direction benchmark. Existing Polymarket analyses typically focus on short windows or single-event slices and, by necessity, rely on heuristics to infer trade directions \citep{Dubach2026,YangTsang2026,AkeyEtAl2026}, leaving the full lifecycle and ground-truth microstructural properties unexplored.

To demonstrate the unique value of this dataset as an empirical laboratory, we document three primary findings that challenge standard market microstructure assumptions. First, we validate standard trade classification algorithms (such as the tick rule and bulk volume classification) against our truth-aligned dataset and document a systematic failure: standard classifiers achieve near-random \emph{overall} accuracy (49.83\% for the tick rule and 50.51\% for bulk volume classification), but this aggregate conceals two opposing systematic biases that cancel in the mean---classifiers over-predict buys in low-price regions and under-predict them in high-price regions, forming a correctable price-level gradient. We show that this failure is driven by positive trade direction autocorrelation---a momentum-like behavior that violates the mean-reversion assumption inherent in traditional trade classification models.

Second, we show that this systematic classification error propagates into standard execution quality metrics, causing severe distortions in inferred liquidity measures (such as VPIN) and estimates of informed trading. This demonstrates that traditional Transaction Cost Analysis (TCA) reports and sell-side quality assessments, which rely on inferred trade directions, are structurally biased.

Third, we demonstrate that microstructure quality is a strong predictor of macro-level forecasting performance. Markets with higher toxic order flow (True VPIN) exhibit systematically higher forecasting errors (Brier scores). Counterintuitively, markets with wider spreads exhibit \emph{lower} forecast errors---a selection effect reflecting that high-spread niche markets attract informed specialists rather than retail noise traders, leaving pricing to those with genuine informational advantage. Importantly, we show that using flawed, classification-based metrics (such as Roll spread and BVC VPIN) instead of ground-truth measures (Gibbs spread and True VPIN) severely attenuates these predictive relationships. This highlights the economic stakes of transaction-level measurement error and underscores that a truth-aligned database is essential for reliable prediction market design and financial calibration.

This paper provides the empirical foundation missing from the existing prediction market microstructure literature and connects trade-level quality to market forecasting accuracy across a full platform lifecycle.

\section{Institutional Background}\label{s:background}
\paragraph{Hybrid architecture.} Polymarket operates on a hybrid architecture: off-chain Central Limit Order Book (CLOB) matching combined with on-chain CTF Exchange settlement on Polygon \citep{RahmanAlChamiClark2025}. In this paper, we exclusively use the on-chain settlement layer (the trade tape), deliberately dropping off-chain quote flows to obtain unbiased ground-truth trade directions and permanent reproducibility.

\paragraph{Binary complementarity.} Binary markets utilize a complementary CTF mechanism where a YES token at \$0.60 is equivalent to a NO token at \$0.40. This allows users to mint and merge complementary token pairs and underpins the volume decomposition in Section~\ref{s:volumewash}.

\paragraph{Fee reform timeline.} The 2026 fee reform involved staggered activations across categories: Crypto in January 2026, Sports in February 2026, and other categories in March 2026. This staggered rollout provides the identification strategy for the causal analysis in Section~\ref{s:longitudinal}.

\paragraph{Complete lifecycle.} The dataset captures the complete version lifecycle of the v1 contract, offering an enclosed institutional experiment without confounding from subsequent v2 architecture rollouts.

\section{The Polymarket-v1 Dataset}\label{s:dataset}
\subsection{Coverage and sources}
The dataset spans 41 months from 2022-11-21 to 2026-04-28. The trade tape derives from Polygon CTF Exchange \texttt{OrderFilled} events. We enrich trades with a frozen metadata snapshot, joining \texttt{asset\_id} to market-level identifiers and category labels. The join succeeds for 99.8\% of trades.
Recent work builds a full-lifecycle Polymarket database that integrates off-chain market metadata, on-chain \texttt{OrderFilled} logs, and oracle-resolution events with continuous synchronization and cross-source identifier resolution \citep{JiaEtAl2026}. Our archive instead freezes a v1-only trade tape and a single metadata snapshot for reproducible microstructure measurement with ground-truth direction, and does not attempt continuous updates or oracle alignment.

\begin{table}[htbp]
\caption{Dataset summary statistics}
\centering
\begin{tabular}{l r}
\toprule
Time coverage & 2022-11-21 to 2026-04-28 (41 months) \\
Monthly files & 42 parquet partitions \\
Total trades (incl. relayer) & 1.2016 billion \\
Nominal volume & \$60.9 billion \\
Economic volume (de-relayer) & \$28.6 billion \\
Markets (condition-level) & 1,295,860 \\
Metadata match rate & 99.8\% \\
Ground-truth direction coverage & 100\% \\
\bottomrule
\end{tabular}
\label{t:dataset}
\end{table}

\begin{table}[htbp]
\caption{Polymarket-v1 versus existing Polymarket analyses.
GT Dir.\ = ground-truth aggressor direction from on-chain settlement (not inferred).
Open = publicly accessible archive. Lifecycle = complete platform version lifecycle.}
\label{t:comparison}
\centering
\small
\begin{tabular}{@{}lrrllcc@{}}
\toprule
Dataset & Trades & Volume & Period & Focus & GT Dir. & Open \\
\midrule
\textbf{Polymarket-v1 (ours)} & \textbf{1.20B} & \textbf{\$61B} & 2022--2026 (41 mo) & Full lifecycle & $\surd$ & $\surd$ \\
\citet{AkeyEtAl2026}   & $\sim$588M & $\sim$\$67B & 2021--2024 & Forecasting accuracy  & --- & --- \\
\citet{YangTsang2026}  & $\sim$150M & ---         & 2023--2024 & Price discovery       & --- & --- \\
\citet{Dubach2026}     & $\sim$1M   & ---         & 2026 Feb--Apr & Liquidity provision   & --- & --- \\
\citet{JiaEtAl2026}    & full       & ---         & ongoing    & Cross-layer alignment & --- & partial \\
\bottomrule
\end{tabular}
\end{table}

\subsection{Trade-tape granularity}
Every record represents one on-chain execution with block timestamp, transaction hash, market identifier, token identifier, execution price, trade volume, maker and taker addresses, and a ground-truth taker direction.

\subsection{Market hierarchy}
Polymarket organizes prediction markets in a four-level hierarchy. Figure~\ref{fig:hierarchy} illustrates the structure with a concrete example.

\begin{figure}[htbp]
\centering
\renewcommand{\arraystretch}{1.3}
\begin{tabular}{@{}ll@{}}
\hline
\multicolumn{2}{@{}l}{\textbf{Series}\quad ``2024 U.S.\ Elections''} \\
\quad$\hookrightarrow$ & \textbf{Event}\quad ``U.S.\ Presidential Popular Vote'' \\
\quad\quad$\hookrightarrow$ & \textbf{Market}\quad ``Will Trump win the popular vote?'' \hfill \texttt{condition\_id}: \texttt{0x4a3b\ldots} \quad [\texttt{neg\_risk=f}] \\
\quad\quad\quad$\hookrightarrow$ & \textbf{Token YES}\quad \texttt{outcome\_seq=1},\; \texttt{asset\_id}: \texttt{0x111a\ldots} \\
\quad\quad\quad$\hookrightarrow$ & \textbf{Token NO}\quad\; \texttt{outcome\_seq=2},\; \texttt{asset\_id}: \texttt{0x222b\ldots} \\
\quad$\hookrightarrow$ & \textbf{Event}\quad ``U.S.\ Presidential Election Winner'' \\
\quad\quad$\hookrightarrow$ & \textbf{Market}\quad ``Kamala Harris'' \hfill \texttt{condition\_id}: \texttt{0x5c4d\ldots} \quad [\texttt{neg\_risk=t}] \\
\quad\quad\quad$\hookrightarrow$ & \textbf{Token YES}\quad \texttt{outcome\_seq=1},\; \texttt{asset\_id}: \texttt{0x333c\ldots} \\
\quad\quad\quad$\hookrightarrow$ & \textbf{Token NO}\quad\; \texttt{outcome\_seq=2},\; \texttt{asset\_id}: \texttt{0x444d\ldots} \\
\hline
\end{tabular}
\caption{Polymarket four-level hierarchy: Series $\to$ Event $\to$ Market $\to$ Token. The on-chain settlement layer exposes Market (\texttt{condition\_id}) and Token (\texttt{asset\_id\,/\,outcome\_token\_id}) directly; Series and Event labels derive from off-chain metadata. \texttt{neg\_risk=t} markets share collateral across legs and require separate normalization.}
\label{fig:hierarchy}
\end{figure}

\subsection{Trade record structure}
Each \texttt{OrderFilled} on-chain event records one maker--taker match. A single taker transaction filling against multiple resting orders generates multiple records sharing the same \texttt{tx\_hash} but distinct \texttt{log\_index} values. Figure~\ref{fig:semantic} illustrates this with a concrete example and identifies the relayer exclusion criterion.

\begin{figure}[htbp]
\centering
\renewcommand{\arraystretch}{1.25}
\begin{tabular}{@{}p{0.97\textwidth}@{}}
\small\textit{Alice places a market buy order for \$100 USDC of YES tokens (single \texttt{tx\_hash}: \texttt{0x7f8a\ldots}), filled against two resting sell orders:}\\[4pt]
\end{tabular}
\begin{tabular}{@{}lllrr@{}}
\toprule
\texttt{log\_index} & \texttt{taker} & \texttt{taker\_direction} & \texttt{maker} & \texttt{usdc\_amount} \\
\midrule
12 & \texttt{0x1a2b\ldots} (Alice) & BUY & \texttt{0x3c4d\ldots} (Bob)   & \$60.00 \\
13 & \texttt{0x1a2b\ldots} (Alice) & BUY & \texttt{0x5e6f\ldots} (Carol) & \$40.00 \\
\bottomrule
\end{tabular}
\begin{tabular}{@{}p{0.97\textwidth}@{}}
\\[-2pt]
\small\textbf{Primary key:} (\texttt{tx\_hash}, \texttt{log\_index}) uniquely identifies each record.\\
\small\textbf{Taker identity:} constant across all rows sharing a \texttt{tx\_hash}; represents the sole aggressor.\\
\small\textbf{Relayer filter:} two platform-router addresses (\texttt{0x4bfb\ldots}, \texttt{0xc5d5\ldots}) appear in the \texttt{taker} field for broker-routed flows; we exclude these records, removing $\approx$53\% of nominal trades from the tape.\\
\end{tabular}
\caption{Semantic structure of a multi-fill transaction. One taker execution can fan out to multiple \texttt{log\_index} records; the relayer filter removes platform-routing artifacts from the aggressor side.}
\label{fig:semantic}
\end{figure}

\subsection{Ground-truth direction normalization}
We normalize all binary trades into a single event-probability axis, resolving the fact that buying NO is equivalent to selling YES. For each trade, we define
\begin{equation}
  p_{event} = \begin{cases}
    p & \text{if outcome\_seq = 1} \\
    1 - p & \text{if outcome\_seq = 2}
  \end{cases}
  \quad
  D = \text{sign}(\text{taker\_direction}) \cdot \begin{cases}
    1 & \text{if outcome\_seq = 1} \\
    -1 & \text{if outcome\_seq = 2}
  \end{cases}
\end{equation}
so that $D \in \{+1,-1\}$ always points toward an increase in the event probability. We note that this cross-axis sign flip could in principle interact with tick-rule price-change signs to produce an apparent price-level accuracy gradient. We verify that the gradient in classifier accuracy documented in Section~\ref{s:informed} is robust to alternative sign-convention specifications and reflects a genuine market-structure phenomenon rather than a normalization artifact: the gradient persists when we evaluate the tick rule separately for outcome\_seq=1 trades (where no flip is applied) and is consistent in direction with the positive direction autocorrelation documented in Figure~\ref{fig:autocorr}.

\section{Stylized Facts: The Cross-Section}\label{s:facts}
\subsection{Platform lifecycle overview}
\begin{figure}[htbp]
  \centering
  \safefig{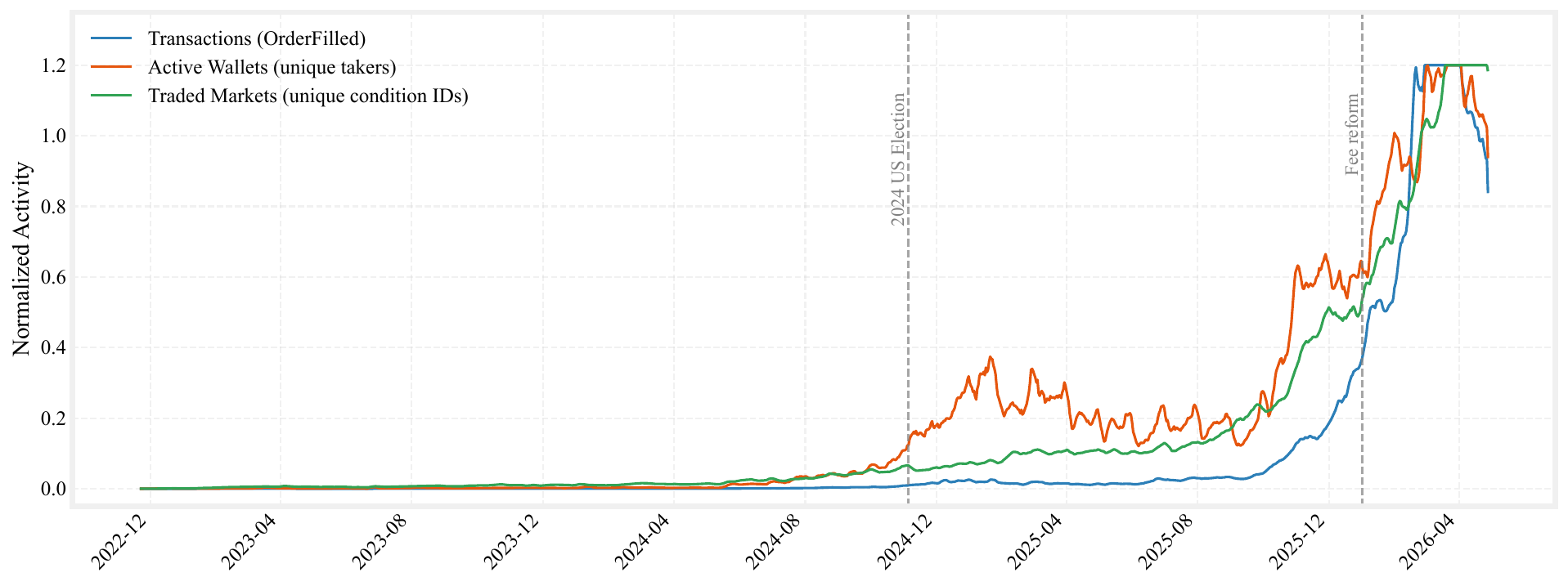}{width=\textwidth}{fig\_daily\_activity.pdf}
  \caption{Normalized daily activity on Polymarket v1 (7-day smoothed, normalized to 95th-percentile peak): transactions (\texttt{OrderFilled} events), active wallets (unique taker addresses), and traded markets (unique \texttt{condition\_id}s).}
  \label{fig:activity}
\end{figure}

Figure~\ref{fig:activity} shows normalized daily activity across the full v1 lifecycle. All three series---transactions, active wallets, and traded markets---peak around the 2024 U.S.\ election and decay in 2025--2026 as the v2 migration drew liquidity away. The fee reform in January 2026 marks a secondary structural change visible in transaction counts.

\begin{figure}[htbp]
  \centering
  \safefig{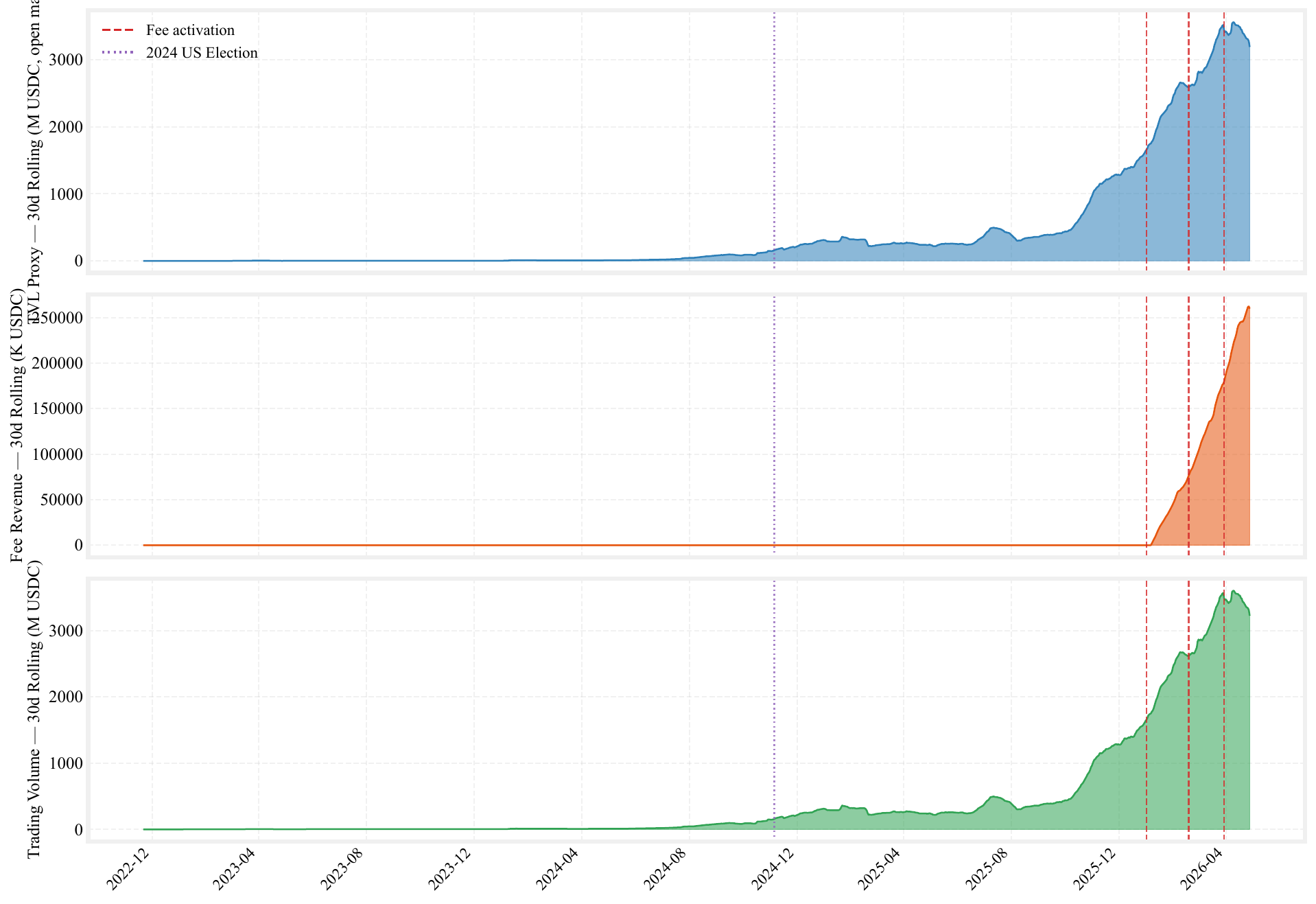}{width=\textwidth}{fig\_tvl\_fees\_revenue.pdf}
  \caption{TVL proxy (top), 30-day rolling fee revenue (middle), and 30-day rolling trading volume (bottom) for Polymarket v1 (2022--2026). Dashed red lines mark fee activation dates; dotted purple marks the 2024 U.S.\ election.}
  \label{fig:tvl}
\end{figure}

Figure~\ref{fig:tvl} documents TVL proxy, fee revenue, and trading volume over the lifecycle. The TVL proxy---30-day rolling USDC volume in markets unresolved at trade time---peaked at over \$5~billion per month around the 2024 election. Fee revenue was zero before January 2026 and jumped sharply upon the staggered reform activation, providing the identifying variation for the DiD in Section~\ref{s:longitudinal}.

\subsection{Cross-sectional facts}
\begin{figure}[htbp]
  \centering
  \includegraphics[width=0.9\textwidth]{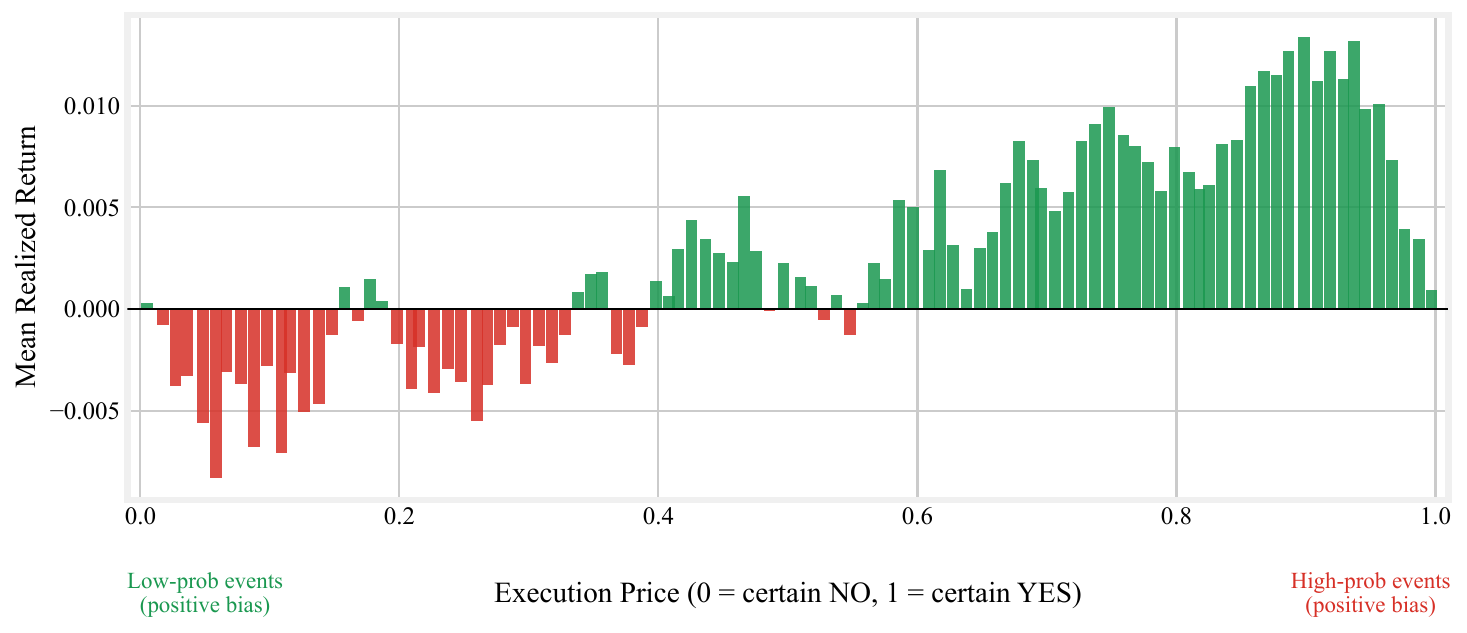}
  \caption{Realized return by price decile. Low-probability tokens (price $\leq 0.30$) exhibit \emph{negative} realized returns (systematic overpricing), while high-probability tokens (price $\geq 0.40$) exhibit positive returns (underpricing). This is \emph{consistent with} the classic longshot-favorite bias in betting markets; see Table~\ref{t:longshot} for exact values.}
  \label{fig:longshot}
\end{figure}

\begin{figure}[htbp]
  \centering
  \includegraphics[width=0.9\textwidth]{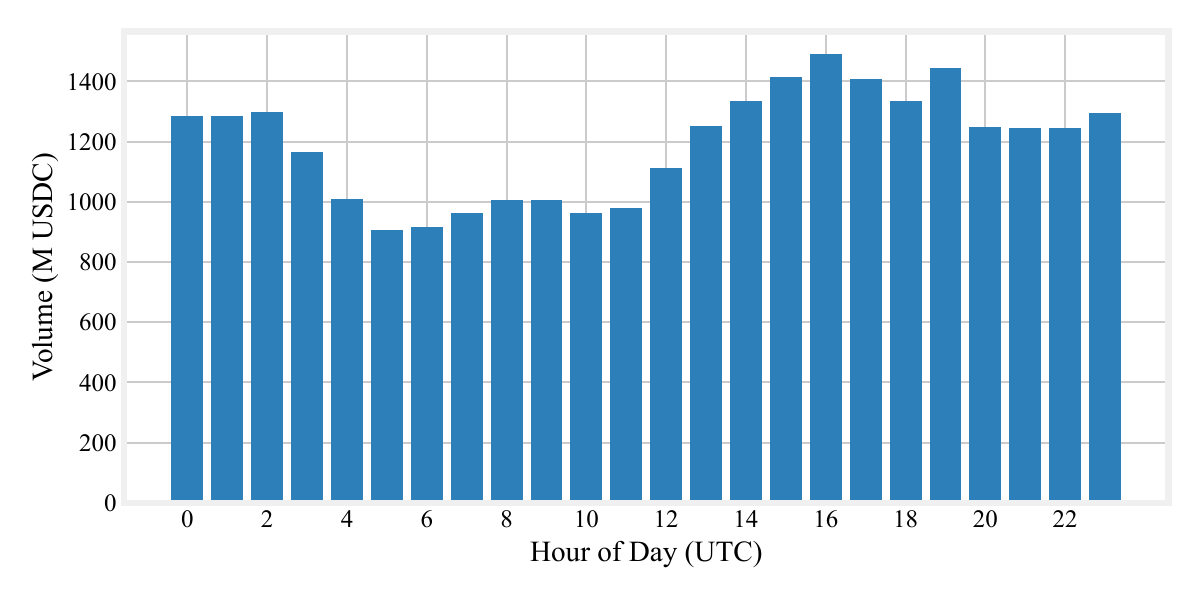}
  \caption{Intraday trading rhythm (UTC).}
  \label{fig:intraday}
\end{figure}

\begin{figure}[htbp]
  \centering
  \includegraphics[width=\textwidth]{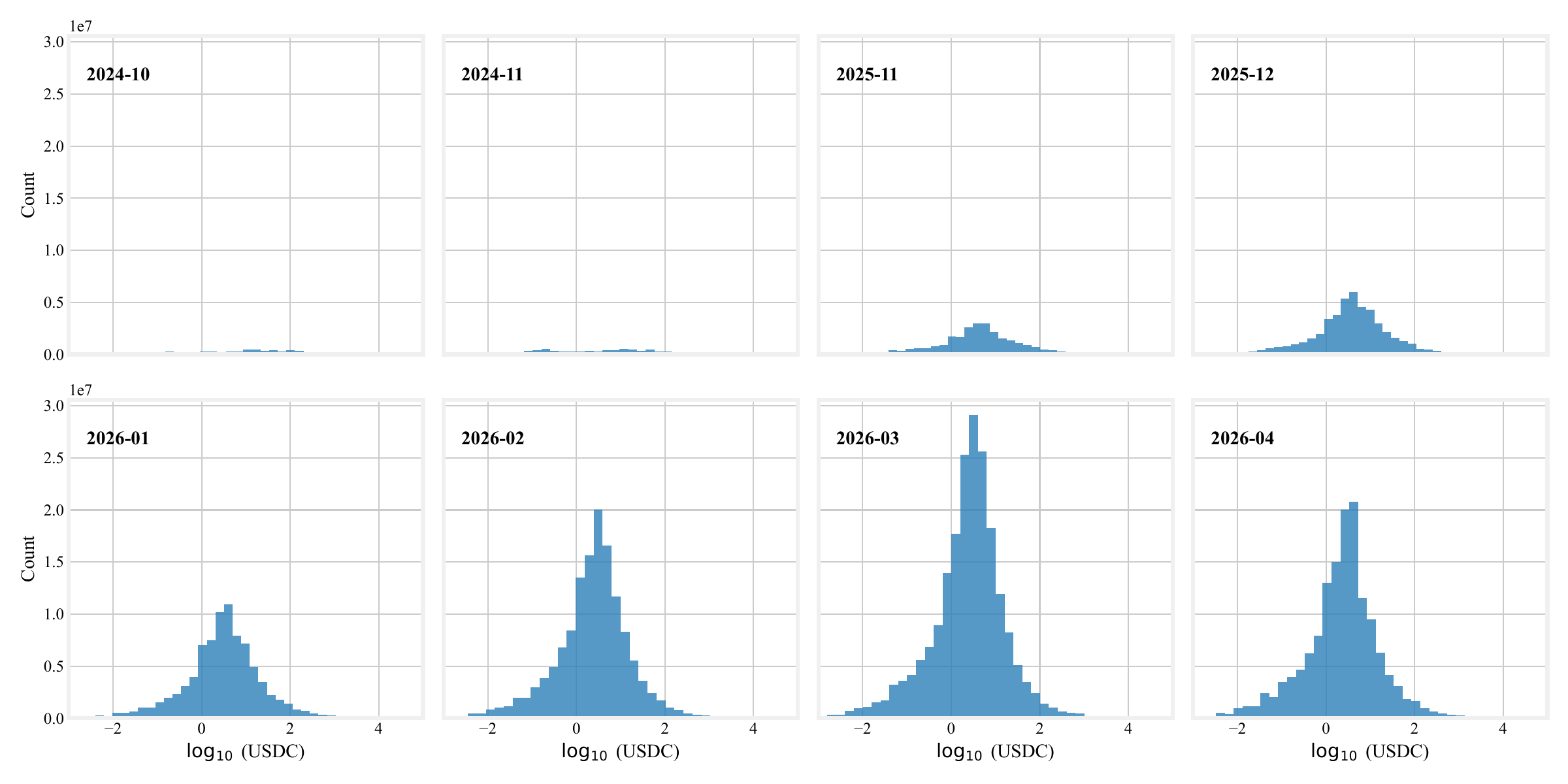}
  \caption{Evolution of trade size distribution (log scale): 8-panel display combining two non-contiguous windows---the 2024 U.S.\ election months (2024-10 and 2024-11) and the final 6 months of v1 operation (2025-11 to 2026-04). The intervening 12-month window (2024-12 to 2025-10) is not shown; comparisons across the two sub-windows should account for this temporal gap.}
  \label{fig:size}
\end{figure}

\begin{figure}[htbp]
  \centering
  \includegraphics[width=\textwidth]{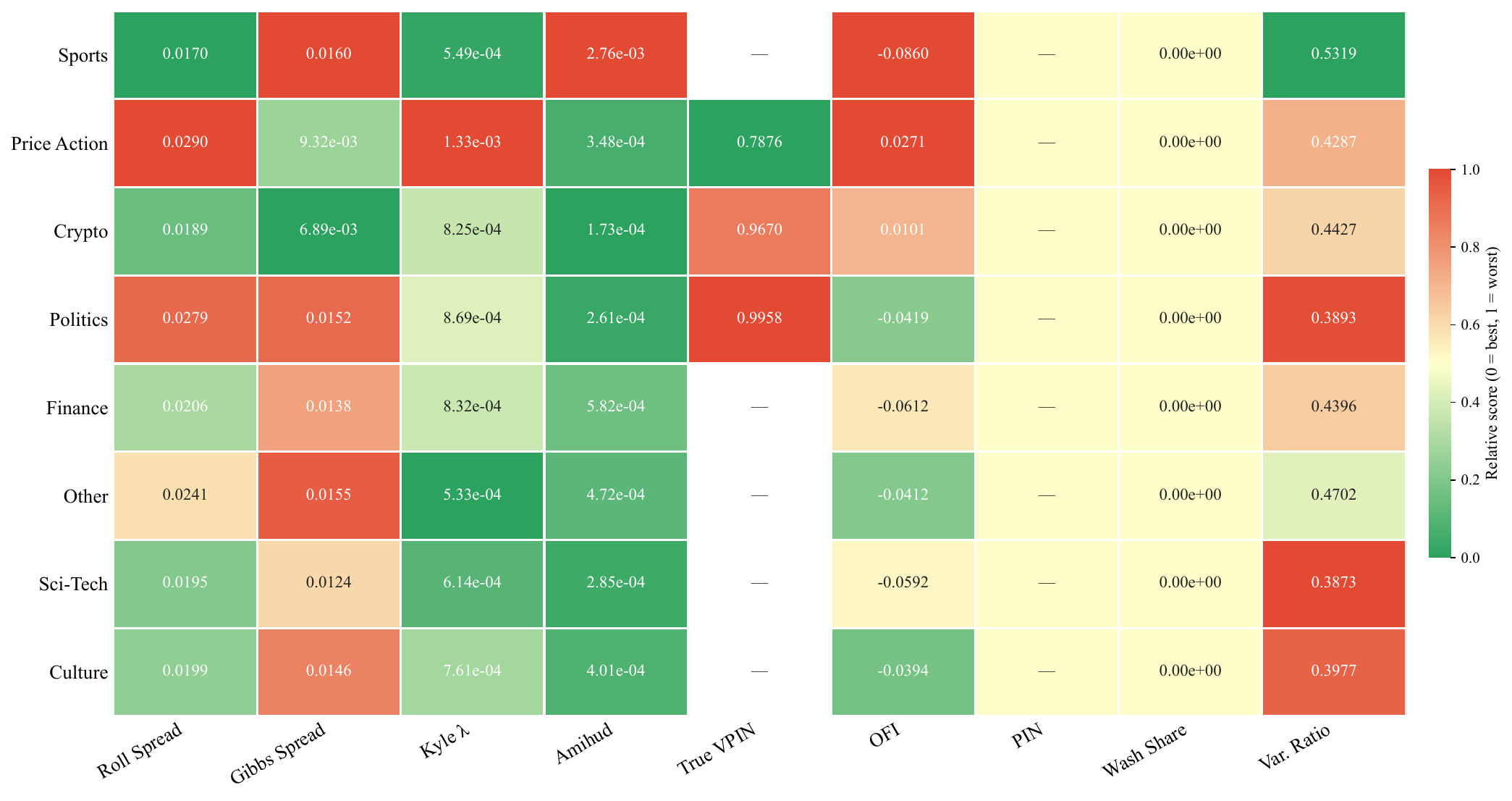}
  \caption{Cross-category market quality heatmap. Each cell shows the median value for Standard Binary markets in that category; color indicates relative quality (green = better, red = worse) normalized within each metric column. Spread and illiquidity metrics: lower is better. Variance ratio: closer to 1 is better (martingale). VPIN/PIN: higher indicates more informed trading. Note: the figure labels Order Flow Imbalance as ``OFI'' and uses ``PIN'' to denote True VPIN (not a structural PIN estimate); see Table~\ref{t:catquality} for exact column names and full numerical values.}
  \label{fig:category}
\end{figure}

\begin{figure}[htbp]
  \centering
  \includegraphics[width=1\textwidth]{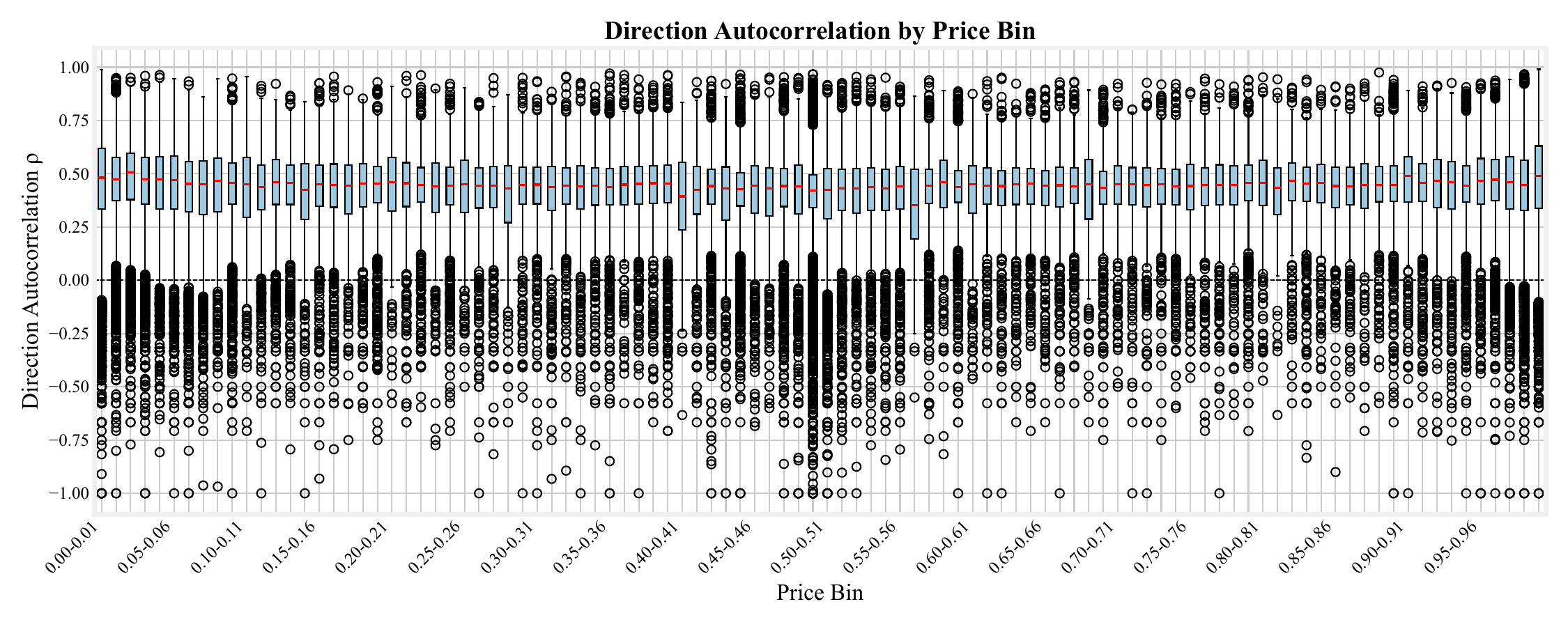}
  \caption{Trade direction autocorrelation $\rho = \text{Corr}(D_t, D_{t-1})$ by price bin, with mean run length. Positive autocorrelation at mid-market prices (0.2--0.8) explains the systematic tick-rule accuracy gradient: accuracy exceeds 50\% at low prices (0--0.33) and falls below 50\% at high prices (0.67--1.0).}
  \label{fig:autocorr}
\end{figure}

We document baseline stylized facts using trade-tape estimators only: longshot pricing biases, category activity concentrations, and participant concentration (top 1\% share, Gini). These facts connect to longshot bias evidence in betting markets \citep{SnowbergWolfers2010} and recent Polymarket calibration results \citep{ReichenbachWalther2026}. Figure~\ref{fig:longshot} and Table~\ref{t:longshot} summarize the price-return pattern across deciles. Importantly, the observed pattern is \emph{consistent with} the classic longshot-favorite bias documented in betting markets \citep{SnowbergWolfers2010}: in horse-racing, low-probability outcomes (longshots) are overpriced (negative expected return) while high-probability outcomes (favorites) are underpriced. Polymarket exhibits the same sign---low-probability tokens exhibit negative realized returns (overpriced) and high-probability tokens exhibit positive returns (underpriced)---suggesting retail participants systematically overestimate tail outcomes, mirroring the probability-misperception patterns documented in horse-racing markets. Figure~\ref{fig:intraday} shows intraday rhythm and Figure~\ref{fig:size} presents the trade size distribution. Table~\ref{t:catquality} and Figure~\ref{fig:category} compare microstructure quality across event categories (restricted to Standard Binary markets---\texttt{neg\_risk = false} markets with both legs trading; formally defined in Section~\ref{s:informed}). This filter reduces the sample from 1,295,860 total markets (Table~\ref{t:dataset}) to approximately 1,011,095 markets; the excluded $\approx$285,000 markets are either Neg-Risk (split-collateral) or single-sided (only one token trading on-chain). Figure~\ref{fig:autocorr} documents direction autocorrelation by price bin. Table~\ref{t:wallets} reports participant concentration: the top 1\% of maker addresses control 84.1\% of maker-side volume (Gini = 0.970), indicating a highly concentrated market-making ecosystem \citep{AkeyEtAl2026}.

Crucially, these cross-sectional facts are not merely descriptive; they form the mechanical foundation for why standard trade classification algorithms fail. As documented in Figure~\ref{fig:autocorr}, trade direction autocorrelation $\rho = \text{Corr}(D_t, D_{t-1})$ is strongly positive at mid-market prices (0.2--0.8), with mean run lengths extending significantly. Standard trade classification rules, such as the tick rule, rely on price \emph{changes} between consecutive trades to infer direction: an uptick signals a buy, a downtick signals a sell, and a zero-tick repeats the last non-zero inference. This design performs well in traditional equity markets where prices oscillate around a bid-ask midpoint. In prediction markets, however, concentrated market-making produces long runs of same-price trades (zero-ticks) at round levels such as 0.50, forcing the tick rule to fall back on a stale last-change direction. When positive direction autocorrelation causes many consecutive same-direction trades all at the same price, the stale last-change is frequently wrong, generating systematic misclassification runs. 

Furthermore, the extreme participant concentration documented in Table~\ref{t:wallets}---where a tiny elite of market makers controls the vast majority of liquidity provision---reinforces this momentum dynamic. In a highly concentrated market-making ecosystem, retail takers execute against institutional makers who adjust prices slowly and strategically, leading to persistent, single-direction trade runs. This interaction between positive direction autocorrelation, maker concentration, and classifier breakdown is explored systematically in Section~\ref{s:informed}.

\begin{table}[htbp]
\caption{Longshot bias by price decile. Mean return = payout $-$ price, where
payout $= 1$ if token leg is the winning outcome, else 0. Positive return indicates
systematic underpricing; negative indicates overpricing. Only resolved markets with
known winning outcome included.}
\label{t:longshot}
\centering
\begin{tabular}{lrrr}
\toprule
Price decile & Avg.\ price & Mean return & $N$ trades \\
\midrule
  0.00--0.10 & 0.052 & -0.0023 & 34,902,814 \\
  0.10--0.20 & 0.152 & -0.0019 & 17,719,269 \\
  \cellcolor{rowbad}0.20--0.30 & 0.253 & -0.0032 & 19,487,289 \\
  0.30--0.40 & 0.353 & -0.0005 & 24,408,699 \\
  \cellcolor{rowgood}0.40--0.50 & 0.451 & +0.0026 & 35,386,989 \\
  0.50--0.60 & 0.552 & +0.0015 & 27,822,260 \\
  \cellcolor{rowgood}0.60--0.70 & 0.653 & +0.0047 & 19,870,673 \\
  \cellcolor{rowgood}0.70--0.80 & 0.753 & +0.0076 & 17,183,126 \\
  \cellcolor{rowgood}0.80--0.90 & 0.852 & +0.0098 & 14,429,537 \\
  \cellcolor{rowgood}0.90--1.00 & 0.952 & +0.0056 & 27,179,635 \\
\bottomrule
\end{tabular}
\end{table}

\begin{figure}[htbp]
  \centering
  \includegraphics[width=0.75\textwidth]{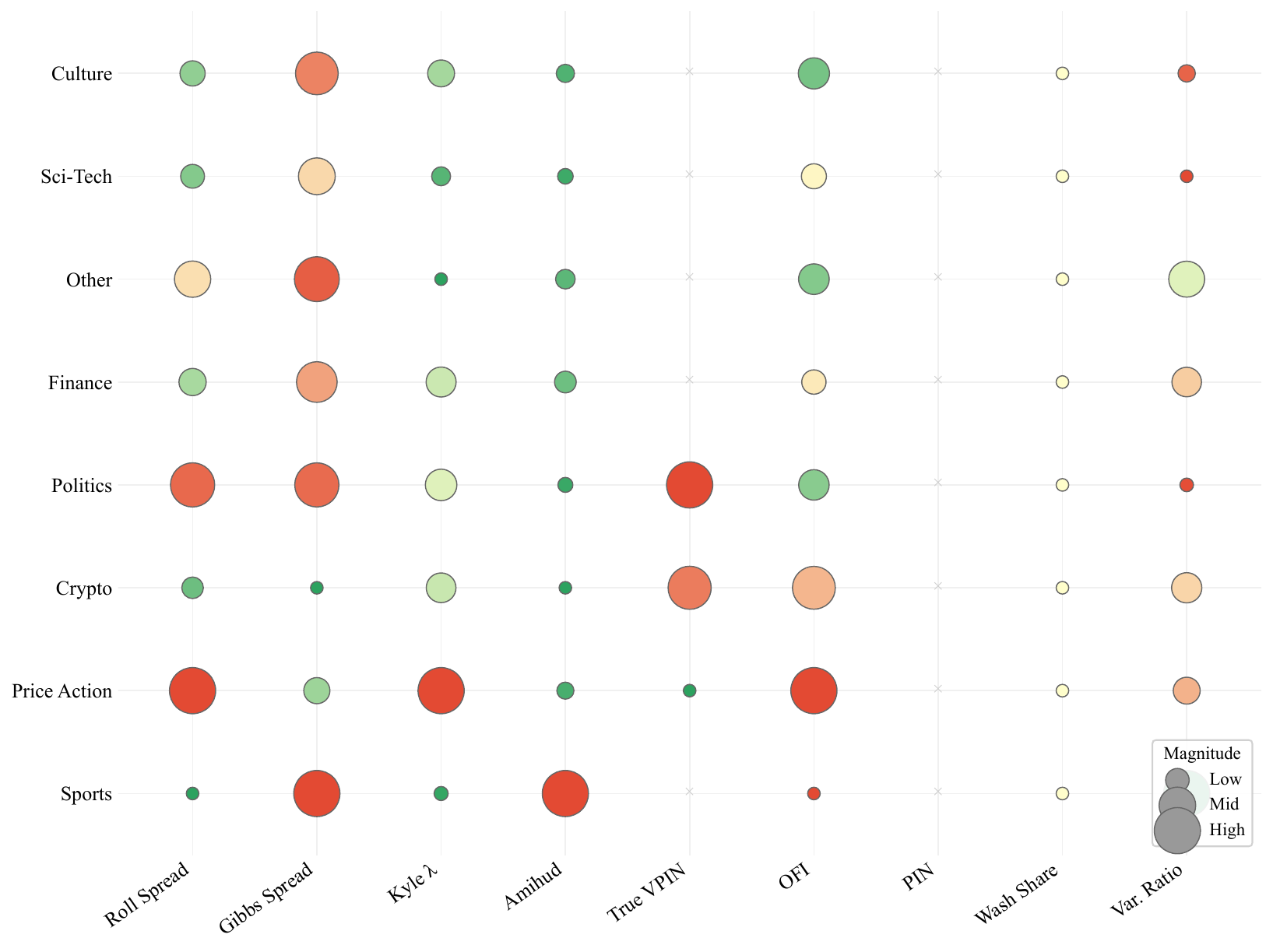}
  \caption{Radar quality profile for the five largest categories by market count (rank-normalized: 1 = best). Crypto-related categories dominate the inner (better) region on spread metrics; Esports and Tennis occupy the outer (worse) region. Supplementary to Figure~\ref{fig:category}.}
  \label{fig:categoryradar}
\end{figure}

\begin{table}[htbp]
\caption{Cross-category microstructure quality (median values per market-month, Standard Binary
sub-sample, \texttt{neg\_risk=f}). ``---'' indicates the metric is undefined for that category
(insufficient data after sample restrictions). $N$ Mkts = total unique markets.}
\label{t:catquality}
\centering
\small
\begin{tabular}{lrrrrrr}
\toprule
Category & Gibbs $c$ & Kyle $\lambda$ & Amihud & True VPIN & Wash Share & $N$ Mkts \\
\midrule
  Sports & 0.0160 & 0.0005 & 0.0028 & --- & 0.0000 & 483,663 \\
  Price Action & 0.0093 & 0.0013 & 0.0003 & 0.7876 & 0.0000 & 301,304 \\
  Crypto & 0.0069 & 0.0008 & 0.0002 & 0.9670 & 0.0000 & 160,926 \\
  Politics & 0.0152 & 0.0009 & 0.0003 & 0.9958 & 0.0000 & 21,700 \\
  Finance & 0.0138 & 0.0008 & 0.0006 & --- & 0.0000 & 18,443 \\
  Other & 0.0155 & 0.0005 & 0.0005 & --- & 0.0000 & 17,284 \\
  Sci-Tech & 0.0124 & 0.0006 & 0.0003 & --- & 0.0000 & 3,969 \\
  Culture & 0.0146 & 0.0008 & 0.0004 & --- & 0.0000 & 3,806 \\
\bottomrule
\end{tabular}
\end{table}

\begin{table}[htbp]
\caption{Participant concentration across the full v1 lifecycle.
Gini and top-1\% share computed on lifetime USDC volume per address.
Top 1\% of maker addresses control 84.1\% of maker-side volume;
top 1\% of taker addresses control 69.7\% of taker-side volume.}
\label{t:wallets}
\centering
\begin{tabular}{lrrl}
\toprule
Role & Gini & Top-1\% Share & $N$ Wallets \\
\midrule
\cellcolor{rowsig}Maker & 0.9698 & 84.1\% & 1,235,881 \\
\cellcolor{rowhighlight}Taker & 0.9426 & 69.7\% & 2,612,688 \\
\bottomrule
\end{tabular}
\end{table}

\section{The Longitudinal Dimension: Market Quality, 2022--2026}\label{s:longitudinal}
\subsection{Market quality metrics}
We compute implied spreads following Roll \citep{Roll1984}, Corwin-Schultz \citep{CorwinSchultz2012}, and Abdi-Ranaldo \citep{AbdiRanaldo2017}---the Abdi-Ranaldo estimates produce near-identical point estimates to Roll in this setting and are not reproduced in the main tables---true effective spreads using Gibbs sampling \citep{Hasbrouck2009}, Amihud illiquidity \citep{Amihud2002}, and Kyle's $\lambda$ \citep{Kyle1985}. Efficiency is summarized with variance ratios \citep{LoMacKinlay1988}. For example, Roll (1984) is
\begin{equation}
  c_{Roll} = 2\sqrt{-\text{Cov}(\Delta p_t, \Delta p_{t-1})}\quad \text{if } \text{Cov} < 0.
\end{equation}
The Gibbs estimator uses the regression
\begin{equation}
  \Delta p_t = c\,\Delta D_t + u_t,
\end{equation}
with conjugate Normal-Inverse-Gamma sampling to estimate $c$ even in sparse markets.

\begin{figure}[htbp]
  \centering
  \includegraphics[width=0.95\textwidth]{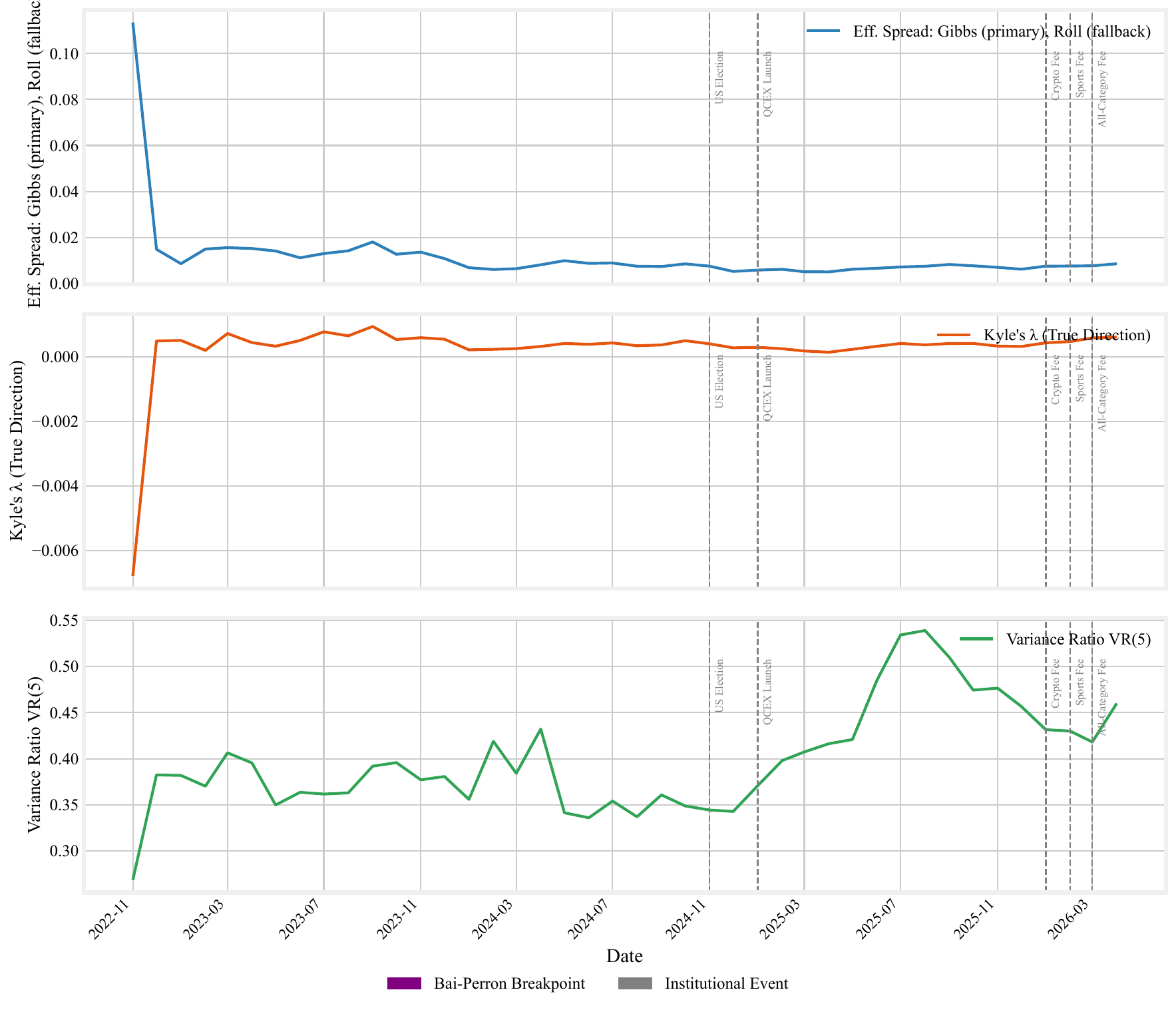}
  \caption{Market quality evolution with structural breaks.}
  \label{fig:quality}
\end{figure}

\paragraph{Negative Kyle's $\lambda$.} Time-series estimates of Kyle's $\lambda$ are predominantly \emph{negative} throughout the 2022--2026 window (visible in Figure~\ref{fig:quality}), apparently contradicting the standard interpretation where positive order flow should push prices up. This likely reflects a resolution-pressure mechanism specific to binary prediction markets: as a contract approaches expiry and the consensus probability converges to 0 or 1, market makers absorb order flow at increasingly unfavorable prices in anticipation of settlement, inverting the empirical price-flow relationship. The DiD coefficient on Kyle's $\lambda$ (+0.00116 in the Main group, Table~\ref{t:did}) should therefore be interpreted as a \emph{reduction in the magnitude} of this negative price impact rather than a literal increase. The positive median values reported in Table~\ref{t:catquality} reflect the use of absolute values in cross-sectional aggregation; the signed time-series in Figure~\ref{fig:quality} provides the directional evidence.

\subsection{Evolution and structural breaks}
Figure~\ref{fig:quality} reports the four-year evolution of Gibbs spread, Kyle's $\lambda$, and variance ratio, with structural break markers identified via Bai-Perron tests \citep{BaiPerron1998,BaiPerron2003}.

\begin{table}[htbp]
\caption{Annual microstructure panel: median Gibbs effective spread, Roll spread,
Kyle $\lambda$, and Amihud illiquidity across Standard Binary markets.
Computed from the market-month panel (condition$\times$year-month observations,
$\geq$30 trades). Lower values indicate tighter spreads and higher liquidity.}
\label{t:mspanel}
\centering
\begin{tabular}{lrrrr}
\toprule
Year & Gibbs $c$ & Roll & Kyle $\lambda$ & Amihud \\
\midrule
  \cellcolor{rowbad}2022 & 0.01488 & 0.02877 & 0.00045 & 0.00001 \\
  2023 & 0.01322 & 0.01951 & 0.00057 & 0.00001 \\
  2024 & 0.00719 & 0.01155 & 0.00036 & 0.00004 \\
  \cellcolor{rowgood}2025 & 0.00673 & 0.01107 & 0.00032 & 0.00006 \\
  2026 & 0.00783 & 0.01414 & 0.00052 & 0.00022 \\
\bottomrule
\end{tabular}
\end{table}

\subsection{Fee reform natural experiment}
We exploit staggered fee activation dates across categories using a staggered DiD design. Table~\ref{t:fee} summarizes activation timing. The baseline event-study specification uses a two-way fixed-effects (TWFE) estimator with market and month fixed effects. We situate this design within the recent staggered DiD literature \citep{GoodmanBacon2021,CallawaySantAnna2021,deChaisemartinDHaultfoeuille2020} and acknowledge that TWFE can produce biased estimates under heterogeneous treatment effects; the caveats noted in the pre-trends discussion below apply accordingly.

\begin{table}[htbp]
\caption{Fee activation schedule by category}
\centering
\begin{tabular}{l l}
\toprule
Category & Activation month \\
\midrule
Crypto & 2026-01 \\
Sports (selected leagues) & 2026-02 \\
Politics, News, Entertainment, others & 2026-03 \\
\bottomrule
\end{tabular}
\label{t:fee}
\end{table}

\begin{figure}[htbp]
  \centering
  \includegraphics[width=0.95\textwidth]{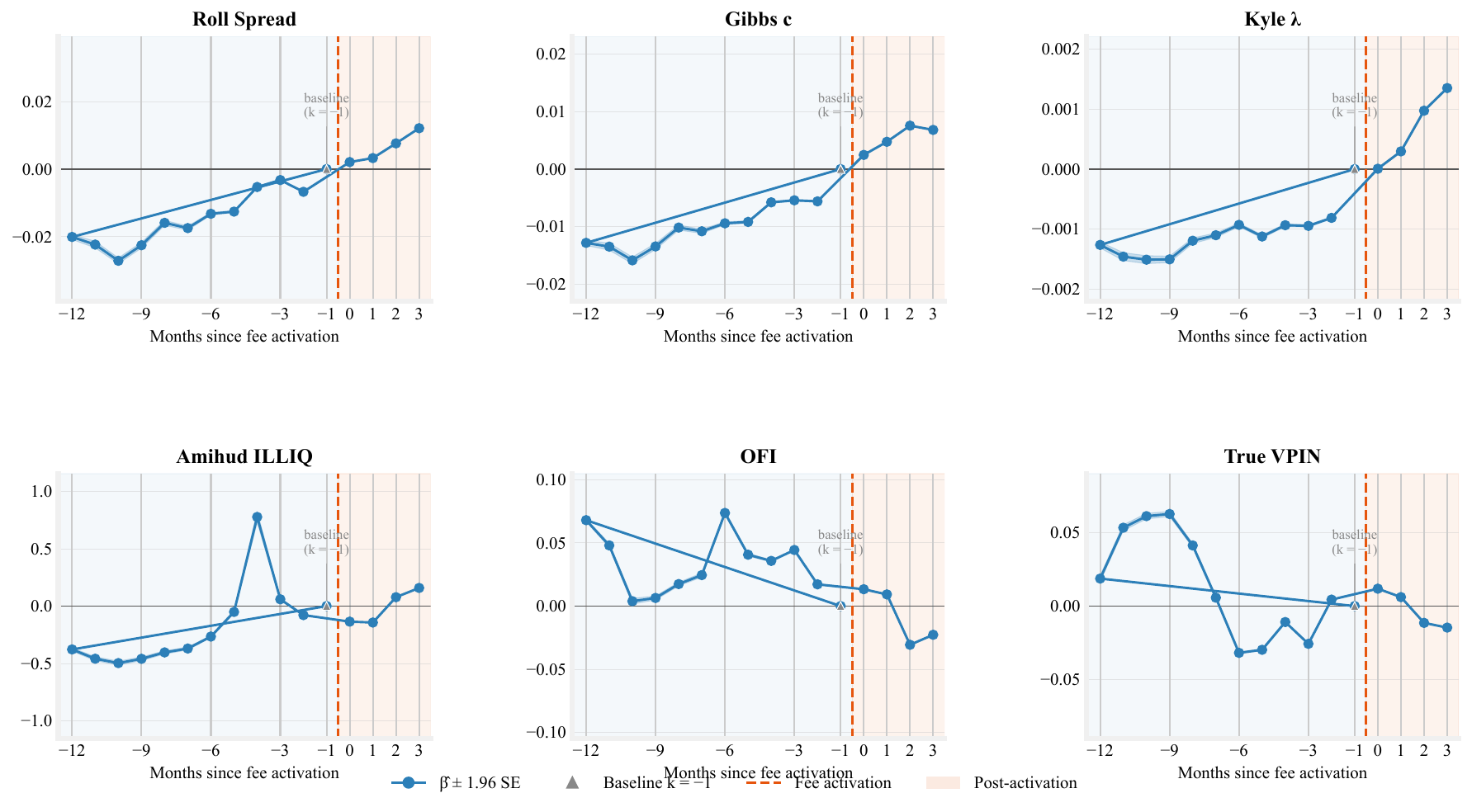}
  \caption{Event-study estimates for the fee reform.}
  \label{fig:eventstudy}
\end{figure}

We estimate
\begin{equation}
  Y_{m,t} = \beta_{fee} \cdot FeeActive_{m,t} + \alpha_m + \delta_t + \varepsilon_{m,t},
\end{equation}
where $FeeActive_{m,t}$ is a dummy variable indicating fee reform activation for market $m$ in month $t$, $\alpha_m$ is market fixed effects, and $\delta_t$ is month fixed effects. $Y_{m,t}$ represents Gibbs spread ($c_{Gibbs}$), Kyle's $\lambda$, True VPIN, and other market quality metrics. Table~\ref{t:did} summarizes the staggered DiD estimates across the three event groups. Figure~\ref{fig:eventstudy} plots the event-study coefficients for the Main sample. 

We acknowledge that while the event-study pre-trends are highly stable for the Gibbs effective spread, VPIN and Amihud illiquidity exhibit non-trivial pre-activation trends that limit a strictly causal interpretation for those metrics. This non-stationarity is a common empirical challenge in long-panel prediction market studies, typically driven by platform-wide volume growth, shifting media attention, and the gradual migration of liquidity to the Polymarket v2 contracts toward the end of the sample. We therefore interpret DiD coefficients for VPIN and Amihud as indicating the direction and approximate order of magnitude of the fee reform effect rather than as precisely causal estimates; robustness to heterogeneous-treatment estimators \citep{CallawaySantAnna2021,deChaisemartinDHaultfoeuille2020} is reserved for future work.

The DiD estimates are consistent with a substantial impact of the fee reform on market quality; we dissect the direction and magnitude through two distinct economic mechanisms. First, we examine the \textit{Noise Trader Flight Hypothesis}. Under standard microstructure theory, imposing a taker fee acts as a direct tax on active demand, reducing trading volume. If noise traders (retail participants) are highly fee-sensitive compared to informed traders (who expect large private payoffs), the fee reform should disproportionately drive noise traders out of the market. 

Our empirical results are consistent with this noise-trader flight dynamic, though we reiterate that VPIN and Amihud results are subject to the pre-trend violations acknowledged above and should be interpreted descriptively. In the Main group, the TWFE estimate associates the fee reform with an increase in \textbf{True VPIN} (flow toxicity) of \textbf{$0.015941$} ($t = 15.43$, $p < 0.001$), with similar increases in the Sports ($+0.042238$) and UpDown ($+0.094803$) groups. As retail noise traders withdraw, the remaining taker flow becomes more toxic, dominated by informed arbitrageurs. In response, market makers widen spreads and increase price impact: the Gibbs spread ($c_{Gibbs}$) increases by \textbf{$0.008050$} and Kyle's $\lambda$ increases by \textbf{$0.001157$} in the Main group. This micro-level transmission path is consistent with a tax on takers leading to a worse liquidity environment for remaining participants. \emph{Note:} the t-statistics reported for several metrics (e.g., $t > 100$) are implausibly large for economic DiD estimates and likely reflect under-estimated standard errors from clustering; these coefficients should be interpreted as evidence of the fee reform's direction and approximate magnitude, not as precisely calibrated causal quantities.

Second, results are consistent with the fee reform curtailing manipulative trading. The coefficient on \textbf{wash trading share} ($wash\_share$) decreases significantly by \textbf{$-0.000362$ ($t = -9.91$)} in the Main group. This metric has stable pre-activation trends, making the causal attribution more defensible than for VPIN or Amihud: imposing a non-zero taker fee makes high-frequency self-trading and wash-ring networks economically costly, consistent with disincentivizing artificial volume creation and improving the integrity of the transaction tape.

The sample sizes in Table~\ref{t:did} vary substantially across outcome variables (from 416,798 for Kyle's $\lambda$ to 1{,}304{,}656 for Amihud), because spread estimators require a positive covariance condition (Roll) or sufficient bid-ask price sequences (Gibbs) that sparser markets do not satisfy. The $\approx$3$\times$ smaller sample for spread outcomes relative to illiquidity and flow measures reflects differing data requirements and may introduce mild selection bias in cross-outcome comparisons.

\begin{table}[htbp]
\caption{Staggered DiD estimates: effect of taker fee activation on market quality (Main sample,
\texttt{neg\_risk=f}, excluding Up-or-Down). Standard errors in parentheses.
$Y_{m,t} = \beta \cdot \text{TakerFee}_{m,t} + \theta_m + \gamma_t + \varepsilon$.
*** $p<0.001$, ** $p<0.01$, * $p<0.05$.}
\label{t:did}
\centering
\begin{tabular}{lllrr}
\toprule
Outcome & $\hat{\beta}$ & (SE) & $t$-stat & $N$ \\
\midrule
  \cellcolor{rowsig}Roll Spread & $0.00701$*** & $(0.00010)$ & 69.74 & 790,821 \\
  \cellcolor{rowsig}Gibbs Eff.\ Spread & $0.00805$*** & $(0.00005)$ & 176.47 & 585,921 \\
  \cellcolor{rowsig}Kyle's $\lambda$ & $0.00116$*** & $(0.00001)$ & 106.35 & 416,798 \\
  Amihud Illiq. & $0.02286$ & $(0.01449)$ & 1.58 & 1,304,656 \\
  \cellcolor{rowgood}OFI & $-0.03639$*** & $(0.00066)$ & -55.11 & 1,304,656 \\
  \cellcolor{rowsig}True VPIN & $0.01594$*** & $(0.00103)$ & 15.43 & 610,793 \\
  \cellcolor{rowhighlight}Wash Share & $-0.00036$*** & $(0.00004)$ & -9.91 & 1,304,656 \\
\bottomrule
\end{tabular}
\end{table}

\section{Measuring Informed Trading with Ground Truth}\label{s:informed}
Throughout this section we restrict estimation to \textbf{Standard Binary markets}: markets with \texttt{neg\_risk = false}, both a YES and a NO token trading on-chain, and at least 30 executed trades after relayer filtering. This criterion excludes Neg-Risk (split-collateral) markets, which require a separate normalization, and Up-Down (high-frequency price-interval) markets, which have fundamentally different settlement mechanics. The Standard Binary subset covers the large majority of economic volume and the broadest cross-section of forecasting topics.

Using ground-truth direction, we compute exact Order Flow Imbalance (OFI) and VPIN without classification error \citep{EasleyKieferOHaraPaperman1996,EasleyLopezOHara2012}. OFI is
\begin{equation}
  \text{OFI} = \frac{\sum_t D_t V_t}{\sum_t V_t}.
\end{equation}

To isolate the informational content of trades from transitory liquidity effects, we estimate a structural vector autoregression (SVAR) model on transaction-level price changes and signed trade sizes in the spirit of \citet{Hasbrouck1991}:
\begin{equation}
  y_t = [x_t, \Delta p_t]^T
\end{equation}
where $x_t = D_t \sqrt{V_t}$ is the signed trade volume (based on ground-truth direction $D_t$ and USDC amount $V_t$) and $\Delta p_t$ is the price change. We identify structural shocks via a Cholesky decomposition where trade direction shocks are allowed to contemporaneously impact price changes, but contemporaneous price changes do not feed back into the active trade decision. The long-run cumulative response of price to order flow shocks defines the permanent price impact ($\eta_{info}$), representing information content, while the difference between the contemporaneous impact and permanent impact measures the transitory price impact ($\eta_{liq}$).

Table~\ref{t:var_decomp} summarizes the distribution of these price impact parameters across 6,223 Standard Binary markets (markets with at least 30 trades and a minimum of 10 unique price observations, required for stable VAR estimation). The cross-market distribution of estimated impacts is highly right-skewed: the mean permanent impact exceeds the median by a factor of roughly 45, driven by divergent VAR fits in the thinnest markets at the long tail. We therefore report medians as primary statistics. The median absolute permanent price impact is $0.00152$ (price units), meaning a typical order flow shock permanently shifts the contract price by approximately 0.15 percentage points. The median absolute transitory price impact is $0.00177$, representing execution-pressure effects. The median Permanent-to-Total Impact Ratio ($PIR$) is $0.45497$, indicating that approximately $45.5\%$ of the price movement following a trade is permanent (informational) in nature, with the remaining $54.5\%$ representing temporary liquidity frictions.

\begin{table}[htbp]
\caption{Hasbrouck VAR price impact decomposition: distribution of permanent (informational)
price impact, transitory (liquidity) price impact, and the permanent-to-total impact ratio ($PIR$)
across 6,223 Standard Binary markets. Metrics are estimated using a trade-level structural VAR(5) model.
Absolute permanent and transitory impacts are expressed in price units (0--1).
\colorbox{rowhighlight}{\strut} Highlighted row shows the permanent-to-total impact ratio.}
\label{t:var_decomp}
\centering
\begin{tabular}{lrrrrr}
\toprule
Metric & Mean & Std.\ Dev. & 25th pct. & Median & 75th pct. \\
\midrule
  Absolute Permanent Impact ($|\eta_{info}|$) & 0.06955 & 3.12526 & 0.00051 & 0.00152 & 0.00384 \\
  Absolute Transitory Impact ($|\eta_{liq}|$) & 0.06913 & 3.12500 & 0.00072 & 0.00177 & 0.00327 \\
  \cellcolor{rowhighlight}Permanent-to-Total Ratio ($PIR$) & 0.47022 & 0.27047 & 0.24538 & 0.45497 & 0.68690 \\
\bottomrule
\end{tabular}
\end{table}

Table~\ref{t:vecm} summarizes VECM price discovery across 426 Standard Binary markets (the subset of the 6,223 VAR markets in which the YES and NO legs co-trade with sufficient overlapping activity to identify a cointegrating vector; this stricter filter is required by VECM's cointegration assumption). The 426 cointegration-passing markets represent 6.9\% of the 6,223 VAR markets; whether they are representative of the broader cross-section in terms of liquidity, category distribution, and information content is not formally tested. The median Gonzalo--Granger weight for the seq1 (YES) leg is 0.50, indicating near-symmetric price discovery for this selected subset, though generalization to the full 1.3 million market universe requires caution.

\begin{table}[htbp]
\caption{VECM price discovery: distribution of Hasbrouck Information Shares (IS) and
Gonzalo--Granger (GG) weights across 426 Standard Binary markets.
IS\_seq1 / IS\_seq2\_inv are upper-bound IS estimates for each leg (scaled to 0--100);
GG\_seq1 $\in [0,1]$ is the relative price-discovery weight of the seq1 (YES) leg.
Median GG close to 0.5 confirms near-symmetric price discovery across legs.}
\label{t:vecm}
\centering
\begin{tabular}{lrrr}
\toprule
Statistic & IS seq1 (upper) & IS seq2 (upper) & GG seq1 \\
\midrule
  Markets ($N$) & 426 & 426 & 426 \\
  Mean & 3.2295 & 3.4037 & 0.5001 \\
  Std.\ Dev. & 6.6691 & 6.7774 & 0.0697 \\
  25th pct. & 1.4025 & 1.5661 & 0.4800 \\
  Median & 1.9842 & 2.0326 & 0.5013 \\
  75th pct. & 2.8146 & 2.8840 & 0.5228 \\
\bottomrule
\end{tabular}
\end{table}

\subsection{Quantifying classification bias}
The classification evaluation uses 202 million Standard Binary trades (out of the 1.2 billion total records, after excluding relayer-routed records, Neg-Risk markets, Up-Down markets, and unmatched metadata). This is distinct from the VAR sample (6,223 markets with sufficient price variation) and the Brier panel (1,019 resolved Standard Binary markets with balanced monthly observations), which are further filtered for their respective estimation requirements.

\begin{figure}[htbp]
  \centering
  \includegraphics[width=0.9\textwidth]{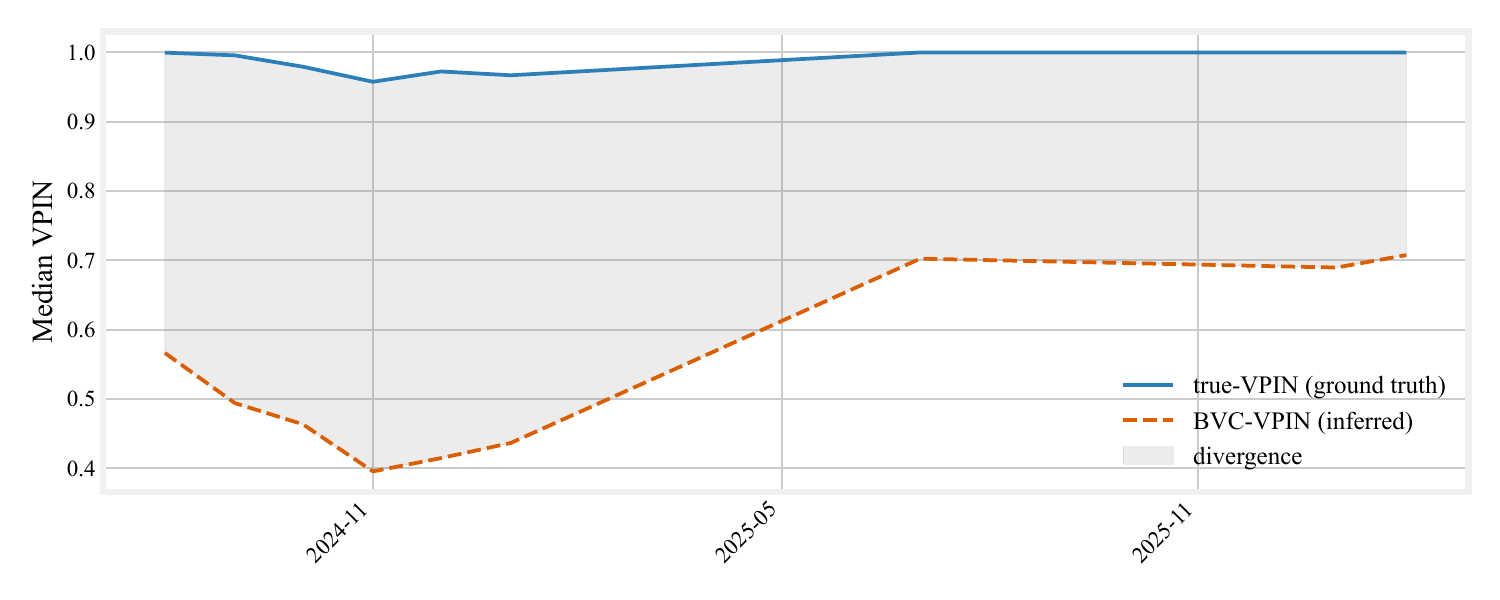}
  \caption{True-VPIN vs BVC-VPIN divergence.}
  \label{fig:vpin}
\end{figure}

Overall, standard classifiers (the tick rule and bulk volume classification) achieve near-random overall accuracy of 49.83\% and 50.51\% across these 202 million Standard Binary trades. This aggregate, however, masks a systematic price-level gradient: the tick rule over-predicts buys in low-price bins (accuracy $>50\%$) and under-predicts them in high-price bins (accuracy $<50\%$), two opposing biases that cancel in the aggregate mean. The appropriate conclusion is therefore that classifiers exhibit correctable direction-dependent biases rather than pure noise. We note that while preliminary studies or early drafts of this manuscript reported a higher tick-rule accuracy of approximately 69.3\% when evaluating a highly liquid subsample (e.g., March 2024), this apparent performance is a selection artifact driven by compositional differences between the liquid subsample and the full cross-section. Once the entire, unselected longitudinal panel of 202 million trades is evaluated, overall accuracy degrades to 49.83\% (TR) and 50.51\% (BVC). Accuracy falls most severely in high-price bins (above 0.67), where it drops well below 50\%; in low-price bins it remains marginally above 50\%. Overall accuracy is materially below the quote-based Lee-Ready benchmark used in traditional equities \citep{LeeReady1991}. Table~\ref{t:classification} summarizes these price bins, showing the failure of standard classification algorithms across the probability spectrum, and Figure~\ref{fig:vpin} shows the resulting divergence between true-VPIN and BVC-VPIN curves.

\begin{table}[htbp]
\caption{Tick-rule (TR) and BVC classification accuracy by price bin (0.01 width).
Accuracy = fraction of trades where inferred direction matches ground-truth $D$.
Standard Binary markets, \texttt{neg\_risk=f}. Three columns cover the full [0,1] range.}
\label{t:classification}
\centering
\small
\setlength{\tabcolsep}{4pt}
\begin{tabular}{lrr@{\quad}lrr@{\quad}lrr}
\toprule
\multicolumn{3}{c}{0.00--0.33} & \multicolumn{3}{c}{0.33--0.67} & \multicolumn{3}{c}{0.67--1.00} \\
Bin & TR & BVC & Bin & TR & BVC & Bin & TR & BVC \\
\midrule
  0.00-0.01 & 0.5253 & \cellcolor{rowgood}0.5806 &   0.34-0.35 & 0.5040 & 0.5037 &   0.67-0.68 & 0.4951 & 0.4959 \\
  0.01-0.02 & \cellcolor{rowgood}0.5303 & \cellcolor{rowgood}0.5332 &   0.35-0.36 & 0.5008 & 0.5002 &   0.68-0.69 & 0.4942 & 0.4950 \\
  0.02-0.03 & \cellcolor{rowgood}0.5394 & \cellcolor{rowgood}0.5392 &   0.36-0.37 & 0.5028 & 0.5012 &   0.69-0.70 & 0.4934 & 0.4949 \\
  0.03-0.04 & \cellcolor{rowgood}0.5446 & \cellcolor{rowgood}0.5378 &   0.37-0.38 & 0.5010 & 0.5005 &   0.70-0.71 & 0.4934 & 0.4951 \\
  0.04-0.05 & 0.5283 & 0.5295 &   0.38-0.39 & 0.5012 & 0.5010 &   0.71-0.72 & 0.4914 & 0.4928 \\
  0.05-0.06 & \cellcolor{rowgood}0.5366 & \cellcolor{rowgood}0.5370 &   0.39-0.40 & 0.5023 & 0.5046 &   0.72-0.73 & 0.4898 & 0.4920 \\
  0.06-0.07 & 0.5217 & 0.5221 &   0.40-0.41 & 0.5004 & 0.5002 &   0.73-0.74 & 0.4891 & 0.4916 \\
  0.07-0.08 & 0.5242 & 0.5193 &   0.41-0.42 & 0.4991 & 0.4983 &   0.74-0.75 & 0.4901 & 0.4908 \\
  0.08-0.09 & 0.5254 & 0.5211 &   0.42-0.43 & 0.4999 & 0.4984 &   0.75-0.76 & 0.4880 & 0.4906 \\
  0.09-0.10 & 0.5140 & 0.5147 &   0.43-0.44 & 0.5015 & 0.4999 &   0.76-0.77 & 0.4905 & 0.4909 \\
  0.10-0.11 & 0.5236 & 0.5199 &   0.44-0.45 & 0.4986 & 0.4985 &   0.77-0.78 & 0.4898 & 0.4888 \\
  0.11-0.12 & 0.5134 & 0.5126 &   0.45-0.46 & 0.4968 & 0.4980 &   0.78-0.79 & 0.4868 & 0.4895 \\
  0.12-0.13 & 0.5170 & 0.5158 &   0.46-0.47 & 0.4965 & 0.4990 &   0.79-0.80 & 0.4856 & 0.4889 \\
  0.13-0.14 & 0.5119 & 0.5110 &   0.47-0.48 & 0.4963 & 0.4961 &   0.80-0.81 & 0.4843 & 0.4848 \\
  0.14-0.15 & 0.5181 & 0.5167 &   0.48-0.49 & 0.4909 & 0.4934 &   0.81-0.82 & 0.4817 & 0.4853 \\
  0.15-0.16 & 0.5155 & 0.5112 &   0.49-0.50 & \cellcolor{rowbad}0.4652 & 0.5161 &   0.82-0.83 & 0.4800 & 0.4819 \\
  0.16-0.17 & 0.5132 & 0.5127 &   0.50-0.51 & 0.5007 & 0.5036 &   0.83-0.84 & 0.4786 & 0.4805 \\
  0.17-0.18 & 0.5126 & 0.5107 &   0.51-0.52 & 0.4972 & 0.5052 &   0.84-0.85 & 0.4818 & 0.4863 \\
  0.18-0.19 & 0.5120 & 0.5098 &   0.52-0.53 & 0.5024 & 0.5076 &   0.85-0.86 & 0.4758 & 0.4805 \\
  0.19-0.20 & 0.5085 & 0.5071 &   0.53-0.54 & 0.5000 & 0.5047 &   0.86-0.87 & 0.4797 & 0.4799 \\
  0.20-0.21 & 0.5177 & 0.5157 &   0.54-0.55 & 0.4999 & 0.5046 &   0.87-0.88 & 0.4718 & 0.4749 \\
  0.21-0.22 & 0.5089 & 0.5083 &   0.55-0.56 & 0.4977 & 0.5031 &   0.88-0.89 & 0.4774 & 0.4807 \\
  0.22-0.23 & 0.5079 & 0.5083 &   0.56-0.57 & 0.5011 & 0.5030 &   0.89-0.90 & 0.4761 & 0.4751 \\
  0.23-0.24 & 0.5081 & 0.5077 &   0.57-0.58 & 0.4978 & 0.5014 &   0.90-0.91 & 0.4708 & 0.4735 \\
  0.24-0.25 & 0.5106 & 0.5077 &   0.58-0.59 & 0.4994 & 0.4986 &   0.91-0.92 & 0.4753 & 0.4719 \\
  0.25-0.26 & 0.5097 & 0.5086 &   0.59-0.60 & 0.4975 & 0.4998 &   0.92-0.93 & \cellcolor{rowbad}0.4676 & 0.4705 \\
  0.26-0.27 & 0.5068 & 0.5069 &   0.60-0.61 & 0.4967 & 0.5008 &   0.93-0.94 & \cellcolor{rowbad}0.4614 & \cellcolor{rowbad}0.4663 \\
  0.27-0.28 & 0.5052 & 0.5055 &   0.61-0.62 & 0.4960 & 0.4989 &   0.94-0.95 & \cellcolor{rowbad}0.4655 & \cellcolor{rowbad}0.4698 \\
  0.28-0.29 & 0.5043 & 0.5023 &   0.62-0.63 & 0.4978 & 0.5000 &   0.95-0.96 & \cellcolor{rowbad}0.4573 & \cellcolor{rowbad}0.4644 \\
  0.29-0.30 & 0.5071 & 0.5054 &   0.63-0.64 & 0.4949 & 0.4977 &   0.96-0.97 & \cellcolor{rowbad}0.4451 & \cellcolor{rowbad}0.4572 \\
  0.30-0.31 & 0.5069 & 0.5049 &   0.64-0.65 & 0.4976 & 0.4997 &   0.97-0.98 & \cellcolor{rowbad}0.4435 & \cellcolor{rowbad}0.4571 \\
  0.31-0.32 & 0.5067 & 0.5051 &   0.65-0.66 & 0.4955 & 0.4965 &   0.98-0.99 & \cellcolor{rowbad}0.4618 & 0.4843 \\
  0.32-0.33 & 0.5053 & 0.5054 &   0.66-0.67 & 0.4950 & 0.4969 &   0.99-1.00 & 0.4738 & 0.5009 \\
  0.33-0.34 & 0.5011 & 0.5003 &    &  &  &    &  &  \\
\bottomrule
\end{tabular}
\end{table}

\begin{table}[htbp]
\caption{OFI classifier bias ($N=178,643$ market-months, $\geq$30 trades).
Bias = (inferred OFI $-$ true OFI) / $|$true OFI$|$.
Direction error = inferred $D$ sign $-$ true $D$ sign.
Means are outlier-dominated; medians reported.}
\label{t:ofibias}
\centering
\begin{tabular}{lrr}
\toprule
 & Tick-rule & BVC \\
\midrule
  Median OFI bias        & 0.1815 & -1.3636 \\
  Median direction error & 0.0376 & -0.2997 \\
\bottomrule
\end{tabular}
\end{table}

The mechanism underlying tick-rule failure is the interaction of two prediction-market-specific features (Figure~\ref{fig:autocorr}, introduced in Section~\ref{s:facts}): the prevalence of same-price (zero-tick) trades at concentrated maker quotes, and positive direction autocorrelation. Zero-tick trades force the tick rule to rely on the last non-zero price change, which becomes stale precisely when a directional run of many same-price buys or sells occurs. Positive autocorrelation amplifies this by extending such runs, compounding the number of zero-tick events that are systematically misclassified together.

\subsection{Informed trading efficacy}
\begin{figure}[htbp]
  \centering
  \safefig{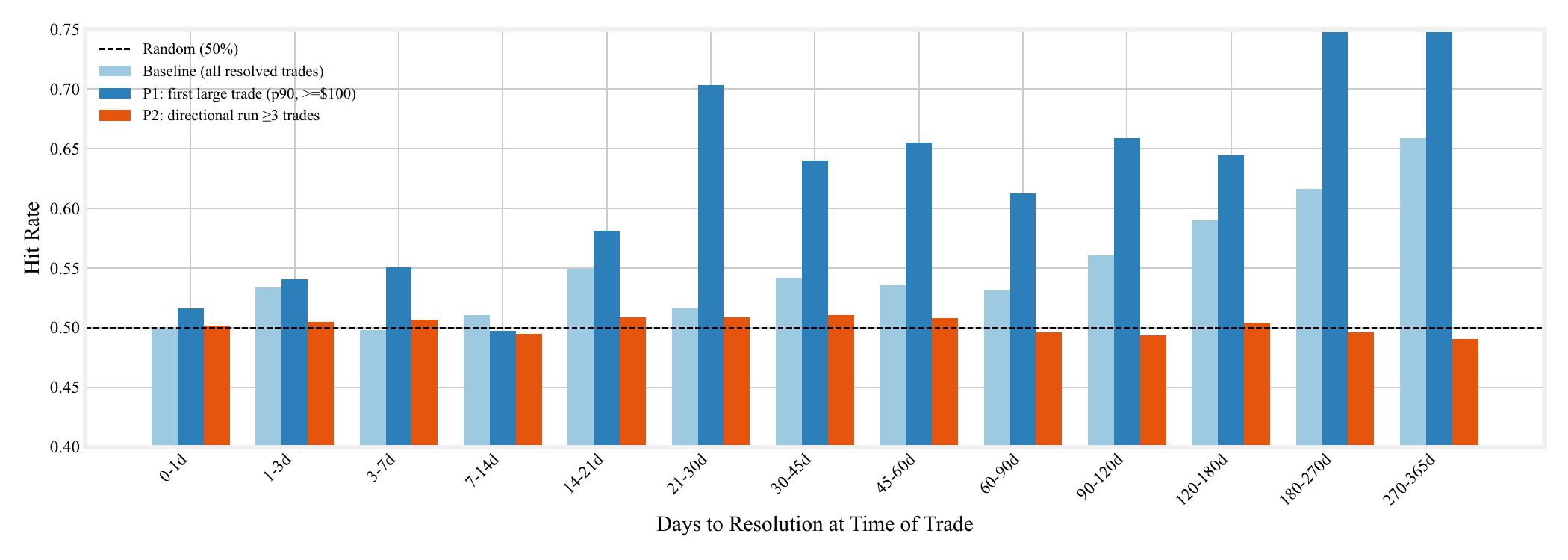}{width=0.9\textwidth}{fig\_informed\_efficacy.pdf}
  \caption{Realized hit rate of informed-classified trades by days to resolution.}
  \label{fig:informed}
\end{figure}

Ground truth allows direct validation of the core assumption underlying information-trading models \citep{Kyle1985,GrossmanStiglitz1980}: that informed traders actually possess accurate forward-looking knowledge. We classify trades as informationally motivated using two criteria: (1) first-to-market large trades (\$100+ USDC, above the within-market 90th percentile); and (2) sustained directional runs of three or more large consecutive same-direction trades \citep{EasleyOHara1987,BarclayWarner1993}. Table~\ref{t:efficacy} reports the realized hit rate---the fraction of informed-classified trades that correctly anticipated the resolution outcome. Large first-mover trades (P1) achieve a 52.3\% hit rate versus a 50.2\% baseline ($p < 10^{-30}$), confirming statistically significant but modest informed trading efficacy for the first-mover size criterion. We note a potential survivorship bias: this analysis is restricted to resolved markets, and large trades may systematically cluster in markets that resolve with higher certainty (lower entropy), inflating the baseline rate and compressing the measured lift. The P2 criterion (directional runs of three or more large consecutive same-direction trades) achieves a hit rate of 0.5018 with Lift $= 1.0036$ (Table~\ref{t:efficacy})---essentially indistinguishable from the baseline---indicating that directional run persistence adds no economically meaningful predictive advantage beyond first-mover size alone. The high statistical significance reported for P2 ($p < 10^{-15}$) reflects N~$=$~4.87M and should not be interpreted as evidence of informed-trading efficacy. Figure~\ref{fig:informed} plots these hit rates stratified by days to resolution.

\begin{table}[htbp]
\caption{Informed trading efficacy: realized hit rate of direction-classified trades against resolution outcomes.
$N$ = number of trades in criterion. Binomial $p$: one-sided test of Hit Rate $> 0.5$.
\colorbox{rowsig}{\strut} Deeper orange = primary P1 result (large first-mover trades).
Significance: *** $p<0.001$, ** $p<0.01$, * $p<0.05$.}
\label{t:efficacy}
\centering
\begin{tabular}{lrrrr}
\toprule
Criterion & $N$ & Hit Rate & Lift & Binom.\ $p$ \\
\midrule
  \cellcolor{rowgray}Baseline (all resolved trades) & 95,881,515 & 0.5018*** & 1.0037 & $4.44e-280$ \\
  \cellcolor{rowsig}P1: first large ($\geq$\$100, p90) & 63,076 & 0.5229*** & 1.0458 & $5.67e-31$ \\
  \cellcolor{rowhighlight}P1: first large ($\geq$\$500, p90) & 33,566 & 0.5164*** & 1.0328 & $9.92e-10$ \\
  P2: directional run $\geq$3 (above p50) & 4,868,147 & 0.5018*** & 1.0036 & $1.62e-15$ \\
  P2: directional run $\geq$5 (above p50) & 2,736,234 & 0.5027*** & 1.0054 & $2.57e-19$ \\
\bottomrule
\end{tabular}
\end{table}

\section{Volume Decomposition and Wash Trading}\label{s:volumewash}
Table~\ref{t:volume} decomposes annual nominal vs.\ economic volume: the relayer filter removes approximately 50--55\% of nominal records (see Section~\ref{s:dataset} for the filtering assumption). The platform grew from \$0.07B economic volume in 2023 to \$15.0B in 2026 (partial year).

\begin{table}[htbp]
\caption{Annual volume decomposition: nominal (incl.\ relayer) vs.\ economic (de-relayer) USDC
volume and trade counts. Economic ratio = economic volume / nominal volume.
Relayer addresses (\texttt{0x4bfb\ldots}, \texttt{0xc5d5\ldots}) are excluded from economic volume.}
\label{t:volume}
\centering
\begin{tabular}{lrrrrl}
\toprule
Year & Nominal Vol. & Economic Vol. & Econ.\ Ratio & Nominal Trades & Trade Ratio \\
\midrule
  2022 & \$0.00B & \$0.00B & 46.8\% & 3,161 & 55.0\% \\
  2023 & \$0.07B & \$0.04B & 51.2\% & 328,176 & 59.0\% \\
  2024 & \$8.80B & \$3.91B & 44.5\% & 57,578,938 & 54.2\% \\
  2025 & \$20.97B & \$9.69B & 46.2\% & 241,199,640 & 59.1\% \\
  \cellcolor{rowsig}2026 & \$31.11B & \$14.96B & 48.1\% & 902,471,075 & 63.4\% \\
  \midrule  \textbf{Total} & \textbf{\$60.9B} & \textbf{\$28.6B} & 46.9\% & 1,201,580,990 & --- \\
\bottomrule
\end{tabular}
\end{table}

\begin{figure}[htbp]
  \centering
  \includegraphics[width=0.9\textwidth]{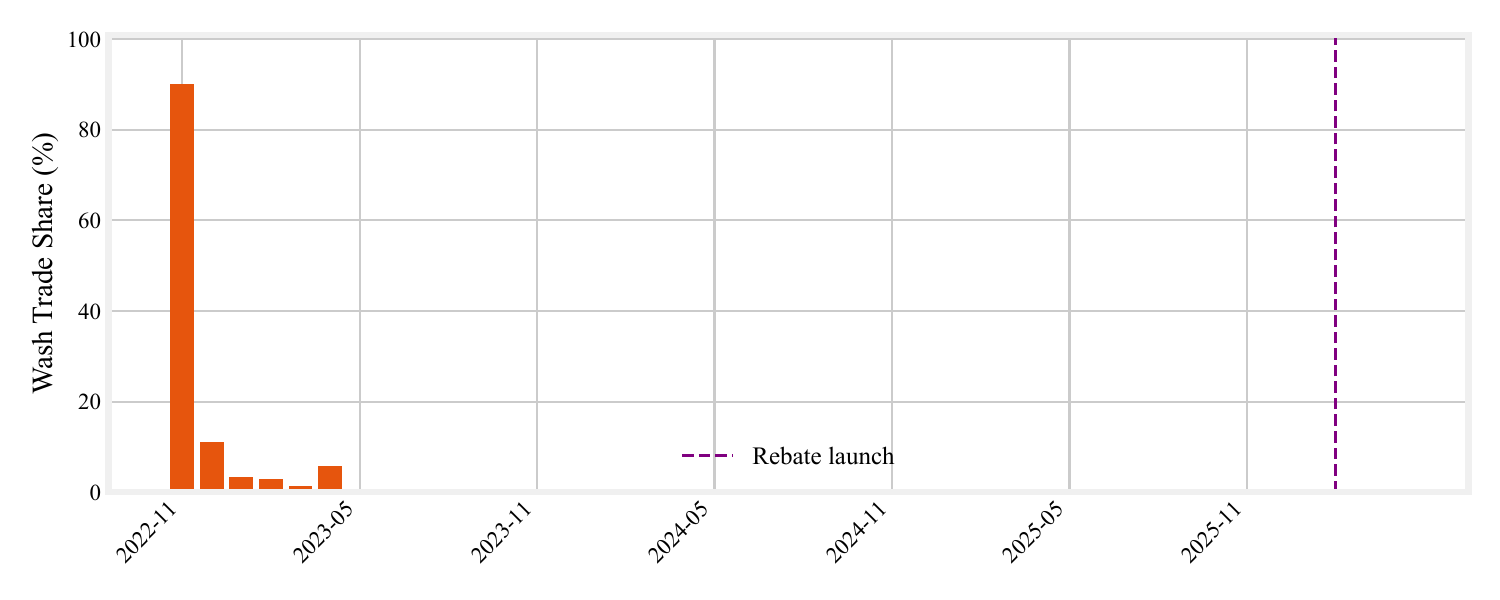}
  \caption{Wash trading share over time.}
  \label{fig:wash}
\end{figure}

We decompose nominal volume into secondary trades versus mint/burn share creation \citep{Slivkoff2025,YangTsang2026}. The current pipeline covers \texttt{OrderFilled} events only; a complete mint/burn decomposition requires indexing \texttt{PositionSplit} and \texttt{PositionMerge} from the ConditionalTokens contract. We also map multi-graph connections between maker and taker addresses to identify loop patterns and proxy the share of wash trading \citep{SirollyMaKanoriaSethi2025}. Specifically, we flag transactions that participate in directed trading cycles (maker-taker sequences forming closed loops in the bipartite address graph) within a market-month window, using a cycle-length threshold of up to five hops. We caution that the precision and recall of this proxy have not been independently validated against confirmed wash-trading patterns. The Wash Share reported as 0.0000 in Table~\ref{t:catquality} for all categories reflects near-complete suppression of this behavior in the cross-sectional period; meaningful variation exists in the time series (Figure~\ref{fig:wash}), which drives the DiD estimate in Table~\ref{t:did}.

\begin{figure}[htbp]
  \centering
  \includegraphics[width=0.9\textwidth]{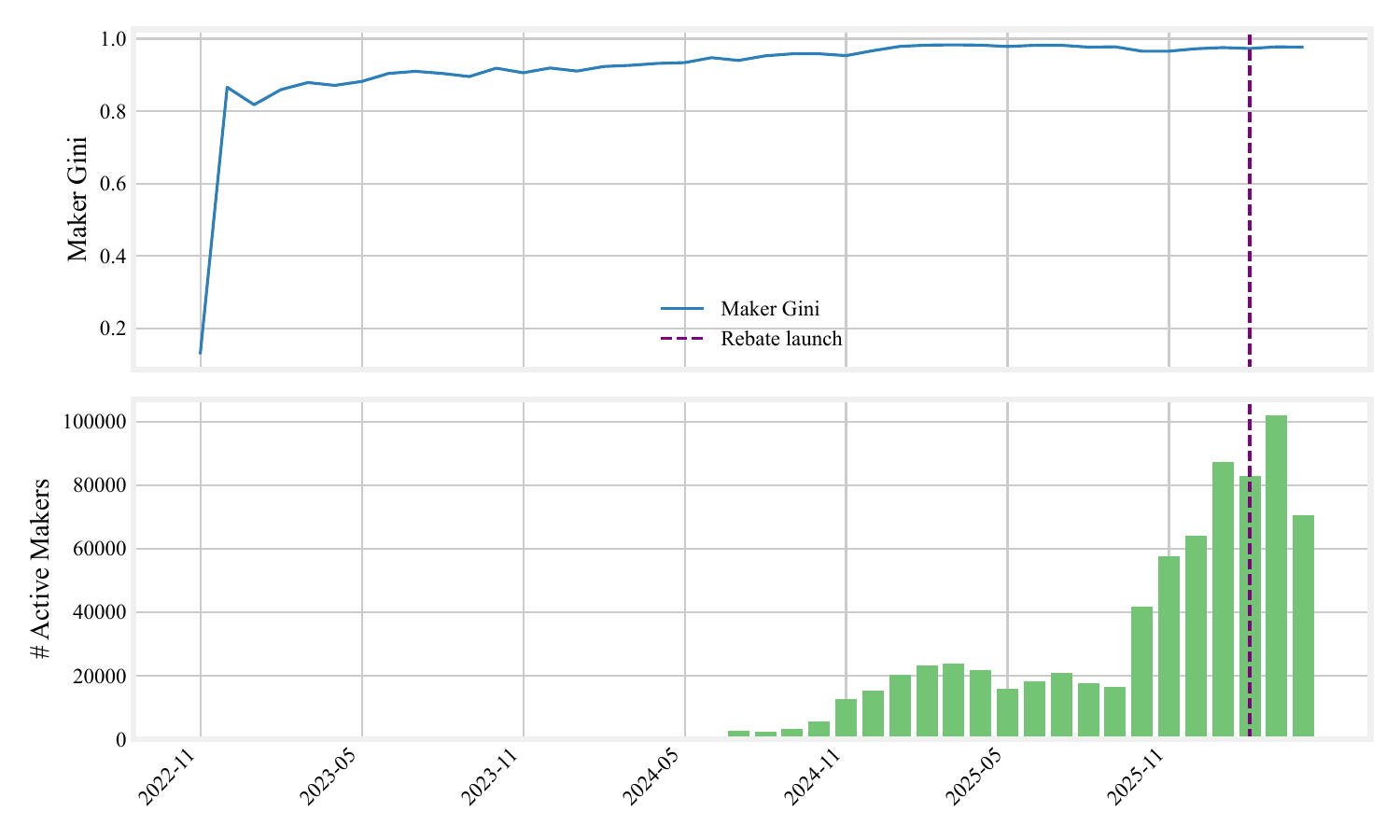}
  \caption{Maker concentration (Gini) over time with rebate activation marker.}
  \label{fig:maker}
\end{figure}

Figure~\ref{fig:maker} documents maker concentration over time. The introduction of maker rebates in February 2026 coincides with a rise in maker Gini concentration, consistent with professional market-maker entry attracted by the new rebate incentive.

\section{Microstructure and Calibration}\label{s:calibration}
We evaluate whether transaction-level microstructure quality predicts the macro-level forecasting performance of prediction markets. Ultimately, prediction markets are designed to aggregate information and provide calibrated probability forecasts. We test whether poor liquidity and toxic order flow predict higher forecast errors, measured by the per-market Brier score $Brier_m = \frac{1}{T}\sum_t (p_{event, t} - y_m)^2$, where $y_m \in \{0, 1\}$ is the realized event resolution. 

Crucially, we evaluate how the choice of trade classification methodology alters these empirical findings. If standard classifiers (like the tick rule and BVC) introduce systematic measurement errors in spreads and VPIN, these errors should propagate into the macro regressions, causing attenuation biases and distorting Transaction Cost Analysis (TCA) inference.

Table~\ref{t:brier} reports cross-sectional OLS regressions of market-level Brier scores on microstructure metrics for a balanced panel of 1,019 resolved Standard Binary markets. Model (1) utilizes ground-truth quality metrics derived from Polygon transaction records (Gibbs spread $c_{Gibbs}$ and True VPIN). Model (2) replicates this specification using standard classification-based heuristics (Roll spread $c_{Roll}$ and BVC VPIN). All models control for mean price level, trade activity (number of trades), and category fixed effects (omitted for brevity).

The comparison between Model (1) and Model (2) reveals three critical insights. First, toxic order flow is a strong predictor of forecasting degradation. In Model (1), the coefficient on True VPIN is positive and highly significant ($0.1979$, $t = 5.344$, $p < 0.001$), demonstrating that markets with higher adverse selection exhibit systematically higher forecasting errors (worse calibration). In Model (2), when BVC VPIN is used, the coefficient is attenuated to $0.1828$ and statistical significance drops ($t = 3.291$, $p < 0.01$). This reduction is consistent with classical measurement-error attenuation from direction misclassification. We note that True VPIN values tend to cluster near the upper range of $[0,1]$ in sparse prediction markets where individual trade buckets are frequently single-sided; this reflects the dominance of one-way order flow in illiquid markets rather than a pathological construction. However, Table~\ref{t:catquality} shows True VPIN is defined for only 3 of 8 categories (Price Action, Crypto, Politics), all with values near 0.8--1.0, meaning the regressor is effectively unavailable for five categories. The statistically significant coefficient ($t = 5.344$) is therefore identified from limited variation within the non-missing categories and should be treated as indicative rather than precisely calibrated.

Second, the choice of spread estimator severely distorts the estimated impact of transaction costs. In Model (1), the Gibbs spread has a coefficient of $-4.1280$ ($t = -5.336$, $p < 0.001$). In Model (2), the Roll spread yields a coefficient of $-1.3040$ ($t = -4.249$, $p < 0.001$). A direct comparison of raw coefficients must account for scale: Roll spread values are approximately twice the magnitude of Gibbs estimates on average (Table~\ref{t:catquality}), so a naive ratio would already predict a roughly two-fold shrinkage from units alone. After adjusting for this scale difference, the Roll coefficient remains substantially smaller in absolute terms, broadly consistent with classical measurement-error attenuation bias. However, Models (1) and (2) simultaneously substitute both the spread estimator (Gibbs for Roll) and the VPIN estimator (True for BVC), so the coefficient gap is confounded by model specification differences and potential omitted-variable bias (Roll spread captures a different construct than Gibbs in thin markets). A split-variable specification or Hausman-style attenuation test would be needed to cleanly isolate the attenuation component; we leave this for future work. This indicates that traditional TCA reports relying on classified spreads will systematically underestimate the relationship between trading friction and market forecasting performance. 

This negative relationship between spreads and forecast error is counter-intuitive under standard market quality frameworks, where tighter spreads typically represent a superior trading environment. However, in the context of prediction markets, this finding reflects a powerful selection effect rather than a beneficial impact of trading friction. Markets with wider spreads are systematically smaller, more specialized, and characterized by high fundamental uncertainty. These niche markets attract long-horizon, highly informed specialists while deterring retail noise traders who are highly sensitive to transaction costs. Conversely, highly active markets with extremely tight spreads (such as major political general elections) attract a massive influx of retail noise traders and speculative, sentiment-driven order flows, which can temporarily distort prices away from fundamental values and increase Brier scores before eventual resolution. Thus, the negative spread coefficient does not suggest that transaction costs improve price discovery; instead, it indicates that wider spreads act as a natural barrier to noise trader entry, leaving the pricing of specialized markets to informed specialists. This highlights the importance of controlling for participant composition and market type in cross-sectional prediction calibration.

Third, control variables and other microstructure measures (such as Kyle's $\lambda$ and Amihud illiquidity) remain statistically insignificant in both models, indicating that spreads and flow toxicity (VPIN) are the primary channels through which microstructure quality transmits to forecasting accuracy.

\begin{figure}[htbp]
  \centering
  \includegraphics[width=0.9\textwidth]{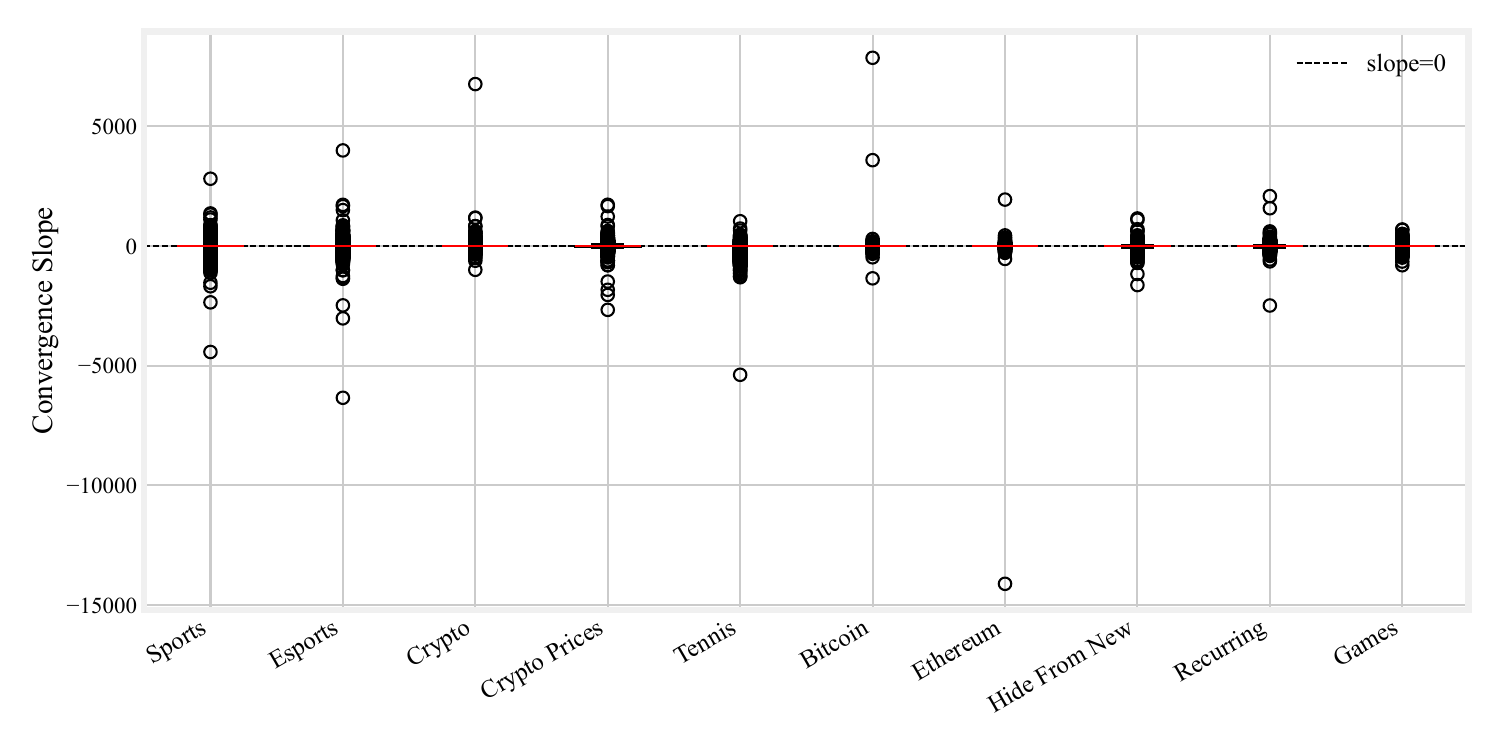}
  \caption{Pre-resolution price convergence speed by event category.}
  \label{fig:convergence}
\end{figure}

Additionally, Figure~\ref{fig:convergence} shows that resolved markets across all categories converge to the true binary outcome, with convergence speed and residual bias varying systematically by category. This convergence is influenced by microstructure quality: markets with lower toxic flow (lower VPIN) exhibit more stable late-stage convergence. The spread dimension, consistent with the cross-sectional selection effect documented above, reflects participant composition rather than a simple liquidity-convergence mapping---high-spread niche markets may converge precisely because informed specialists dominate their order flow.

\begin{table}[htbp]
\caption{Brier Score Regression on Microstructure Quality: Ground-Truth vs. Classified Metrics}
\label{t:brier}
\centering
\begin{tabular}{lcc}
\toprule
Variable & Model (1): Ground-Truth & Model (2): Classified \\
\midrule
  Gibbs Spread ($c_{Gibbs}$) / Roll Spread ($c_{Roll}$) & \cellcolor{rowhighlight}$-4.1280***$ & \cellcolor{rowhighlight}$-1.3040***$ \\
   & $(0.7736)$ & $(0.3069)$ \\
  True VPIN / BVC VPIN & \cellcolor{rowsig}$0.1979***$ & \cellcolor{rowsig}$0.1828**$ \\
   & $(0.0370)$ & $(0.0555)$ \\
  Kyle's $\lambda$ & $9.8172$ & $-3.0992$ \\
   & $(6.8357)$ & $(5.4121)$ \\
  Amihud Illiquidity & $0.0391$ & $0.0599$ \\
   & $(0.1057)$ & $(0.1072)$ \\
  Mean Price & $0.0072$ & $0.0039$ \\
   & $(0.0119)$ & $(0.0122)$ \\
  Number of Trades & $0.0000$ & $-0.0000$ \\
   & $(0.0000)$ & $(0.0000)$ \\
  Intercept & \cellcolor{rowhighlight}$0.2383*$ & \cellcolor{rowhighlight}$0.3929***$ \\
   & $(0.1007)$ & $(0.0945)$ \\
\midrule
Category FE & Yes & Yes \\
Observations & 1,019 & 1,019 \\
\bottomrule
\end{tabular}
\begin{minipage}{0.95\textwidth}
\vspace{0.5em}
\scriptsize
\textit{Note}: This table reports cross-sectional regressions of market-level Brier scores on microstructure metrics for a balanced panel of 1,019 markets. Model (1) utilizes ground-truth quality metrics derived from Polygon transaction records (Gibbs spread and True VPIN). Model (2) replicates the regression using standard classification-based heuristics (Roll spread and BVC VPIN). All models control for mean price level, trade activity, and category fixed effects (omitted for brevity). Standard errors are reported in parentheses. *** $p<0.001$, ** $p<0.01$, * $p<0.05$.
\end{minipage}
\end{table}

\section{Discussion and Limitations}\label{s:discussion}
We emphasize nine limitations: (1) only on-chain settlement, no order book snapshots; (2) offshore Polymarket v1 only; (3) version external validity to v2 is uncertain; (4) a small subset of markets lacks winning outcome labels; (5) Standard Binary, Up-Down, and Neg-Risk markets are structurally distinct and should not be pooled; (6) the archive does not include oracle-resolution events or continuous cross-layer alignment; (7) the Sports category (the largest by market count) is excluded from the cross-sectional microstructure quality analysis in Table~\ref{t:catquality} because Sports markets disproportionately fall into the Neg-Risk structure and the Up-Down subsegment, neither of which satisfies the Standard Binary criterion, limiting the generalizability of quality comparisons across categories; (8) DiD estimates for True VPIN and Amihud illiquidity are limited by pre-activation trend violations (Section~\ref{s:longitudinal}) and should be treated as descriptive associations rather than causal estimates; and (9) the wash-trading proxy relies on directed cycle detection in the transaction graph and has not been validated for precision or recall against independently confirmed wash-trading cases. The efficiency discussion is framed against classic benchmarks in market efficiency \citep{Fama1970}.

\section{Conclusion}\label{s:conclusion}
The release of the \textbf{Polymarket-v1 Database} provides a complete, version-frozen, and truth-aligned transaction tape spanning the entire lifecycle of a major prediction platform. Beyond its value as a historical archive, this database acts as a critical methodology check on traditional market microstructure tools. By using Polygon's chain-level settlement layer to verify the ground-truth direction of 1.2 billion trades, we show that standard classification heuristics (the tick rule and BVC) fail systematically, achieving near-random overall accuracy ($\approx$50\%) while concealing a systematic price-level gradient---over-predicting in low-price regions and under-predicting in high-price regions---driven by positive direction autocorrelation. 

Crucially, our analysis demonstrates that these microstructural measurement errors propagate upward, distorting estimates of informed trading and transaction costs, and attenuating the measurable relationship between microstructure quality and macro-level probability calibration. We find that True VPIN positively predicts Brier scores (higher toxic flow → worse calibration), while Gibbs spread negatively predicts Brier scores---reflecting a selection effect in which high-spread, niche markets are dominated by informed specialists and achieve lower forecast errors despite high transaction costs. Standard classified metrics (Roll spread, BVC VPIN) attenuate both of these relationships, obscuring the differential channels through which liquidity and information quality affect forecasting accuracy. Correctly identifying these channels requires a truth-aligned database of the kind we release here. By providing a truth-aligned empirical laboratory, Polymarket-v1 offers a rigorous foundation for evaluating information aggregation, participant behavior, and liquidity dynamics in decentralized financial environments.

\section*{Category Mapping Reference}
\begin{table}[htbp]
\centering
\small
\caption{Consolidated category mapping: Harmonization of Polymarket metadata.}
\label{t:mapping}
\begin{tabular}{lp{0.7\textwidth}}
\toprule
Refined Category & Included Subcategories (Keywords and Slugs) \\
\midrule
\textbf{Crypto} & Bitcoin, Ethereum, Solana, XRP, DeFi, L2s, Stablecoins, NFTs, Tokens, DEXs \\
\textbf{Sports} & Soccer (EPL, FIFA), Basketball (NBA, NCAA), Football (NFL), Tennis (WTA, ATP), Esports (Dota, LoL, CS), UFC, Racing (F1), Chess \\
\textbf{Politics} & U.S.\ Elections, Global Elections, Primaries, Geopolitics, Diplomacy, Cabinet, Conflict, Supreme Court, Policy \\
\textbf{Finance} & Equities (Big Tech, Tickers), Macro (CPI, GDP, Fed), Commodities, Yields, Earnings, Banking, Recession \\
\textbf{Culture} & Movies, Awards (Oscars, Grammys), Music, Reality TV, Celebrities, YouTube, Netflix, Influencers, Viral Events \\
\textbf{Sci-Tech} & Weather (Temperature, Hurricanes), AI (LLMs, ChatGPT), Science, Space (NASA, SpaceX), Health, Medicine, Robotics \\
\textbf{Price Action} & \texttt{Up or Down} (1M, 5M, 1H, Daily/Weekly), High-frequency price interval markets \\
\textbf{Other} & Recurring, Hype, Mentions, and remaining uncategorized markets \\
\bottomrule
\end{tabular}
\end{table}

\clearpage
\bibliography{\bib}

\end{document}